\documentclass[12pt,tightenlines,eqsecnum,floats,aps,amsmath,amssymb,nofootinbib,prd]{revtex4}

\usepackage{setspace}
\usepackage{graphicx}
\usepackage{xcolor}
\usepackage{amsmath,amssymb,amsfonts,amsthm,mathrsfs}
\usepackage{mathtools}
\usepackage{enumerate} 
\usepackage{braket}
\DeclareMathOperator{\sign}{sgn}

\begin{document}

\title{Canonical analysis of the gravitational description of the $T\bar{T}$ deformation}

\author{Florencia Benítez\footnote{florenciab@fing.edu.uy}, Guzmán Hernández-Chifflet\footnote{guzmanhc@fing.edu.uy} and Esteban Mato\footnote{emato@fing.edu.uy}}
\affiliation{
Instituto de F\'{\i}sica, Facultad de Ingenier\'{\i}a (Universidad de la República), Julio Herrera y Reissig 565, 11300 Montevideo, Uruguay.}

\begin{abstract}
The description of the $T\bar{T}$ deformation in terms of two-dimensional gravity is analyzed from the Hamiltonian point of view, in a manner analogous to the ADM description of general relativity. We find that the Hamiltonian constraints of the theory imply relations between target-space momentum at finite volume which are equivalent to the $T\bar{T}$ finite volume flow equations. This fully-quantum $T\bar{T}$ result emerges already at the classical level within the gravitational theory. We exemplify the analysis for the case when the undeformed sector is a collection of $D-2$ free massless scalars, where it is shown that --somewhat non-trivially-- the target-space two-dimensional Poincar\'e symmetry is extended to $D$ dimensions. The connection between canonical quantization of this constrained Hamiltonian system and previous path integral quantizations is also discussed. We extend our analysis to the ``gravitational'' description of $J\bar{T}$-type deformations, where it is found that the flow equations obtained involve deformations that twist the spatial boundary conditions.
%We begin by considering matter consisting of a single real scalar field whose Lagrangian depends only on Lorentz invariant combinations of the field's first derivatibes, but it is otherwise arbitrary. It is found that, already at the classical level, the mode zero of the constraints imply a relation between target-space energy and momentum (Dirac observables within the canonical formalism) which is analogous to the $T\bar{T}$ finite-volume energy flow equation. 
%We later obtain the analogous equations in the case in which the scalar field is complex. For the special case of a free, real massless scalar field we extend the phase space to include the matter fields and then discuss its relation to a three dimensional string. We then briefly discuss the possibility of coupling to other matter sectors. The passage from the canonical formulation to the path-integral formulation is later discussed in detail.
\end{abstract}
\maketitle

\section{Introduction}
The so-called $T\bar{T}$ deformation has gathered considerable attention in recent years due to its connection to a variety of different aspects of quantum field theory, ranging from integrability to holography and 3D gravity. This deformation, originally introduced in \cite{Smirnov:2016lqw,Cavaglia:2016oda}, consists in a procedure to define a one-parameter family of quantum field theories starting from a seed or ``undeformed'' theory. The procedure may be defined as follows \cite{Smirnov:2016lqw}. Any 2D QFT may be infinitesimally deformed by the $T\bar{T}$ operator, defined as

\begin{equation}
T\bar{T} \equiv \det\left(T_{\mu\nu}\right)
\end{equation}

As shown in \cite{Smirnov:2016lqw}, this deformation is well defined due to the fact that, when regularizing this product of operators by point splitting, any singularities in the coincident limit appear only as total derivatives, and thus are absent from the infinitesimal deformation, which is built from the spacetime integral of this composite operator. This argument relies on the conservation of the stress energy tensor, which holds for any relativistic 2D QFT. Thus any relativistic 2D QFT may be infinitesimally deformed in this manner, so that the deformation defines a vector field in the space of relativistic 2D QFTs. Integral curves of this vector field define one-parameter families of theories, indexed by some parameter $\lambda$. Following some such curve starting from some particular theory $S_0$ --the "seed" or "undeformed" theory-- one obtains at finite $\lambda$ the theory $S_\lambda$, which is often referred to as the $T\bar{T}$ deformation of $S_0$\footnote{In this paragraph the letter $S$ doesn't nececesarily refer to the action but rather more abstractly to some 2d quantum field theory $S$.}.\\    
\indent Interest in these one-parameter families of theories has been due to the fact that they satisfy some remarkable properties. For instance, as noticed first in \cite{Smirnov:2016lqw,Cavaglia:2016oda}, the finite volume spectrum of a given one-parameter family satisfies the exact equation

\begin{equation}
\partial_\lambda E = E\partial_RE + \frac{P^2}{R}
\label{burgers}
\end{equation}   
\noindent which in hydrodynamics is known as Burgers' equation. Here $R$ denotes the size of the compact spatial direction and $P$ is the total spatial momentum, which is conserved under the deformation. This is a remarkable formula: while finite volume spectra of QFTs are in general very difficult to calculate, for $T\bar{T}$-deformed theories these can be obtained by solving the PDE in \eqref{burgers}, assuming the spectrum of the seed theory is known.\\ 
\indent In this work we elaborate on the idea put forward in \cite{dubovsky2017asymptotic,dubovsky2018tt} that the $T\bar{T}$ deformation may be realized by coupling the undeformed theory to a two dimensional gravitational sector. In \cite{dubovsky2017asymptotic} it was shown that this reproduces the $T\bar{T}$ deformed S-matrix, whereas \cite{dubovsky2018tt} showed how the $T\bar{T}$-deformed finite-volume spectrum emerges in the gravitational theory.\\  
\indent The proposal of \cite{dubovsky2018tt} suggests that\footnote{This is a slight reformulation of the action presented in \cite{dubovsky2018tt} to contemplate the possibility of a non flat geometry in target-space (see for instance \cite{Mazenc:2019cfg} for a discussion on this).} 
at finite deformation parameter $\lambda$ the deformed theory may be described by the action

\begin{equation}
S_{\lambda} = S_0[g,\phi] + S_{grav}[X,e]
\label{ttbargravityaction}
\end{equation}

\noindent where

\begin{equation}
S_{grav} = \frac{-1}{2\lambda}\int \epsilon_{ab}(X^*E-e)^a\wedge (X^*E-e)^b
\label{actiongravity}
\end{equation}

\noindent and $S_0[g,\phi]$ is the action of the undeformed theory coupled to the dynamical metric $g$ of the gravitational sector, with $\phi$ standing for the fields in the undeformed theory. The formulation of the gravitational part of the action requires the addition of an auxiliary spacetime with fixed geometry which we will refer to as ``target space''. This auxiliary spacetime is taken to have the same topology as the spacetime in which we are formulating the undeformed theory, which we will refer to as ``worldsheet'', although in general no stringy interpretation will be implied. The fixed geometry of the target space is described by the dyad $E^a$. In the gravitational action the dynamical metric on the worldsheet appears through the dyad $e^a$, although the coupling to matter - that is, to the undeformed action $S_0$- is accomplished through the coupling to the two-dimensional metric $g$ built out of $e^a$ in the standard way. The gravitational action also includes an auxiliary field $X$ which is a map from the worldsheet to target space in a particular homotopy class, on which we will elaborate below for the cases of interest, and $X^*E$ stands for the pullback to the worldhseet of the target-space form $E^a$. Thus the dynamical fields in the gravitational sector are the worldsheet dyad $e^a$ and the map $X$, whereas the dyad $E^a$ is to be thought of as a background field.\\
\indent In this paper we analyse this gravitational theory from the Hamiltonian point of view, in analogy of the Arnowitt-Deser-Misner (ADM) \cite{Arnowitt:1962hi} analysis of classical general relativity (see for instance \cite{Bojowald:2010qpa}). From the canonical point of view, the theory behaves as a constrained Hamiltonian system, with the constraints generating the gauge transformations (diffeomorphisms) of the system when acting on canonical variables. The dynamics is restricted to the constraint surface inside of the phase space of the theory. One of the main results of this paper is to show that, when considering the gravitational theory in finite volume, imposing the aforementioned constraints necessarily requires the observables associated to target-space time translations (the physical energy and momentum) to satisfy the $T\bar{T}$-flow equation \eqref{burgers}. It is interesting that this happens already when imposing the constraints at the classical level. \footnote{A similar analysis to that done here can be found in \cite{Tolley:2019nmm}, although in this paper the focus is more on how the burgers equation for target-space observables emerges from the constraints, and we consider a somewhat different class of undeformed theories to analyze their spectrum, perhaps more general in a certain sense. For related work, see also \cite{de2011resummation,hassan2012ghost,alberte2013reduced,hinterbichler2012interacting,ondo2013complete}}\\
\indent For simplicity, in this paper we will focus on scalar matter with only one-derivative-per-field terms. However, we expect the main framework of our analysis to be applicable to more general matter sectors and our conclusions to extend there as well. \\ 
\indent We also analyze the gravitational description of the $T\bar{T}$ at the quantum level. In \cite{dubovsky2018tt}, a path integral quantization of this theory was introduced, and the torus partition function

\begin{equation}
    Z = \frac{\int De DX}{V_{Diffs}}e^{S_{grav}}Z_0[g(e)]
    \label{torusPartfunCovariant}
\end{equation}

\noindent was calculated, showing that this theory indeed reproduces the $T\bar{T}$-deformed  spectrum given by \eqref{burgers}. Here $Z_0[g]$ is the torus partition function of the undeformed theory with background metric $g(e)$ (built out of $e$) and $V_{Diffs}$ is the volume of the group of spacetime diffeomorphisms, which must be factored out of the integration measure via a Faddeev-Popov type procedure.\\
\indent Our objective here will be to compare with the results of \cite{dubovsky2018tt} starting from the canonical perspective. The starting point of the approach considered here is the construction of the so-called ``reduced'' phase space: the phase space obtained by quotienting the constraint surface by the gauge transformations generated by them. In practice, this phase space is constructed by picking a representative of each gauge orbit by a suitable gauge-fixing condition. By choosing a specific gauge, it is shown that the reduced phase space is identical to that of the undeformed theory, for which we assume a canonical quantization prescription exists. We consider the partition function over the quantized reduced phase space built taking the generators of spacetime translations in the gravitational theory as energy and momentum. We then compare this canonical partition function to that given by \ref{torusPartfunCovariant}, and attempt to rederive the $T\bar{T}$ deformed spectrum. The procedure we follow is straightforward and similar to that followed in many textbooks \cite{Itzykson:1980rh,weinberg1995quantum} to go from the canonical quantization of gauge-fixed gauge theories to the covariant Faddeev-Popov path integral in terms of the action functional. We present this calculation in some detail to focus on the analysis of the path integral measure obtained from this procedure and compare it with that of \cite{dubovsky2018tt}. Since in that work the path integral measure played a crucial role in getting the $T\bar{T}$ result, it is interesting to compare with the measure obtained from the canonical analysis.\\
\indent Regarding the general presentation of our results, many of the ideas applied here are well-known textbook level concepts about the canonical analysis of constrained Hamiltonian systems (e.g. \cite{henneaux}). We spell these out in detail to keep the presentation self-contained and because we believe that this system is simple yet interesting enough to warrant a detailed exposition exemplifying this type of analyses. \\   
\indent The rest of the paper proceeds as follows. In Section \ref{classical}, we develop the canonical formalism for the $T\bar{T}$-gravitational theory at the classical level and obtain our main result regarding the deformed spectrum. In section \ref{sec3massless} we discuss the free massless scalar field case and its connection to non-critical strings. In section \ref{JTsection} we apply similar ideas to a $J\bar{T}$-type of deformation. Section \ref{quantum} discusses the emergence of the covariant path integral quantization of \cite{dubovsky2018tt} from canonical quantization. Finally, we present some conclusions drawn from our analysis and outline future work. 

\section{Canonical analysis of $T\bar{T}$ gravity - classical equations of motion} \label{classical}
We now proceed to study the theory specified in \eqref{ttbargravityaction} in the canonical formalism at the level of the classical equations of motion. As explained in the introduction, this theory is expected to describe the $T\bar{T}$ deformation of the seed theory $S_0$ at deformation parameter $\lambda$. From the point of view of the gravitational interpretation, the undeformed theory $S_0$ is the ``matter'' sector to which the gravitational field $e^a$ couples to. Our objective will be to see how the formula \eqref{burgers} for the finite volume spectrum arises from the point of view of the canonical formalism.\\
\indent We will proceed as follows. We first foliate spacetime into spatial surfaces as required by the canonical formalism and identify the canonical variables and the gravitational and matter hamiltonians. These are seen to be of the ``pure-constraint'' form --as usual in the canonical formalism for gravity-- the constraints being imposed by the equations of motion of the non-dynamical parts of the metric field, which act as Lagrange multipliers. We then identify the target-space observables and via a partial gauge fixing, we will show that the zero modes of the constraints yield relations between these observables equivalent to the $T\bar{T}$ flow equations.\\
\indent As mentioned before, we begin by foliating spacetime by spatial surfaces, as required by the canonical formalis. Since we are interested in placing the theory in finite volume, this implies that we need to consider the theory in a spacetime with the topology of a cylinder $S\times \mathbb{R}$. Notice that the prescription for the gravitational description of the $T\bar{T}$ deformation requires both worldsheet and target-space to be of this form. The worldsheet cylinder will be coordinatized by the pair $(t,x)$, where the compact spatial coordinate $x$ satisfies $x\sim x+1$. We take the target space cylinder to have circumference $R$: this will be the ``physical'' volume our finite-volume spectra will eventually depend on. Target-space recoordinatizations are realized as field redefinitions of the field $X^a$. We choose a coordinatization where we take the target-space dyad to be constant, which without loss of generality we can take to be of the form \footnote{In our notation, greek indices are used for tangent space indices both of the worldsheet and the target space spacetimes. Latin letters are reserved for the frame field internal indices. When specializing the indices to particular values, the spacetime indices will turn into $t$ or $x$, whereas the internal indices take the values $0$ and $1$}

\begin{equation}
E_{\mu}^a = \delta^a_{\mu}.
\end{equation}

\noindent With $E$ constant in our coordinatization of target space, it is useful to define 

\begin{equation}
X^a \equiv X^{\mu} E_{\mu}^a
\end{equation}

\noindent In these terms the action \eqref{ttbargravityaction} becomes

\begin{equation}
S = S_0 + S_{grav} \label{ttbargravityactioncylinder}
\end{equation}

\noindent with

\begin{equation}
S_{grav} = -\frac{1}{2\lambda}\int d^2x \epsilon_{ab}\epsilon^{\alpha\beta}\left(\partial_{\alpha}X^a-e^a_{\alpha}\right)\left(\partial_{\beta}X^b-e^b_{\beta}\right)\label{gravityactioncylinder}
\end{equation}

As mentioned in the introduction, the field $X$ in our gravitational sector must be restricted to a particular homotopy class. In this case, where $X$ describes a mapping between the worldsheet and target space cylinders, we fix $X$ so that the mapping has winding $1$. The target space is coordinatized in such a way that this implies

\begin{align}
X^1(t,x+1) = X^1(t,x) + R  
\end{align}

\noindent In order to implement this it is useful to perform the change of variables 

\begin{equation}
X^a = X^a_w + Y^a,
\end{equation}

\noindent where $X^a_w$ is some fixed function implementing the winding and $Y^a$, which satisfies periodic boundary conditions on the worldsheet spatial circle, is our new dynamical field.\\ 
\indent We now proceed to analyze the theory specified by the action \eqref{ttbargravityactioncylinder} in the canonical formalism. The first step is to identify what the phase space variables are and the symplectic structure in said phase space. After performing the aforementioned split of spacetime, we notice that the action $S_{grav}$ is first order in time derivatives, and that there are no time derivatives acting upon $e^a_t$:

\begin{equation}
S_{grav} = \frac{1}{\lambda}\int dt dx \left(\epsilon_{ab}X^a\partial_t e^b_x -  e^a_t\epsilon_{ab}\left(\partial_x X^b + e^b_x\right) \right) \label{Sgravsplit} 
\end{equation}

\noindent from which we would conclude that the phase space variables in the gravitational sector are $Y^a$ and $e_x^a$, the corresponding symplectic form being given by

\begin{equation}
\Omega_{grav} = - \frac{1}{\lambda}\int dx \epsilon_{ab}\delta X^a \wedge \delta e^b_x
\end{equation}

\noindent From $S_{grav}$ we can conclude that $e^a_t$ is not be a phase space variable but rather plays the role of a Lagrange multiplier. These conclusions may however be altered depending on the nature of the matter action $S_0$. %Indeed, if there are time derivatives acting on the metric in $S_0$, in principle this could turn $e^a_t$ into a dynamical variable and alter the symplectic structure of the theory so that it is not described by $\Omega_{grav}$ in the gravitational sector. 
In order to avoid these complications, we consider a somewhat less general undeformed theory and take $S_0$ to be the action of a single scalar field with only one derivative per field terms, as mentioned in the introduction.\\ %In cases with either scalars with higher derivative terms or fields in other representations of the Lorentz group the symplectic structure is more complicated. In fact, in these cases --in general-- even the symplectic structure of the theory without coupling to gravity has its complications. Therefore, we will restrict to this simplified situation for now and will comment further on the general case in section .\\
\indent The matter action $S_0$ thus may be written in terms of an arbitrary scalar function $K$ of the scalars $\phi$ and $u\equiv g^{\alpha\beta}\partial_{\alpha} \phi\partial_{\beta} \phi$
 
\begin{equation}
S_{0} = \int d^2x\sqrt{-g}K(u,\phi)
\label{scalaraction}
\end{equation} 

Under these circumstances the only phase space variables in the gravitational sector are indeed $e^a_x$ and $X^a$, with the symplectic form on the full phase space being given by

\begin{equation}
\Omega = \Omega_{grav} + \int dx \delta\phi\wedge\delta \pi.
\label{symplecticform}
\end{equation}

\noindent Notice that in our conventions the field $\pi$ --the conjugate momentum to $\phi$--is a spatial one-form. The variables $e^a_t$ are indeed non-dynamical in this case and act as Lagrange multipliers, as will be seen explicitly below.\\ 
\indent The Hamiltonian of the full deformed theory is of the form

\begin{equation}
H =  H_{grav} + H_{mat}
\end{equation}

\noindent where $H_{grav}$ is the contribution from the gravitational sector and $H_{mat}$ denotes the matter sector contribution. From \eqref{Sgravsplit} we see that $H_{grav}$ is given by 

\begin{equation}
H_{grav} = \frac{1}{\lambda}\int dx e^a_t \epsilon_{ab}\left(\partial_x X^b + e^b_x\right)
\label{Hgrav}
\end{equation}

\noindent The matter contribution $H_{mat}$ is obtained from the Legendre transform of \eqref{scalaraction} in the standard way, although with the extra complication that the action \eqref{scalaraction} for the scalar is somewhat general and is in the presence of an arbitrary metric. Details on the procedure to obtain $H_{mat}$ are provided in Appendix \ref{appscalarham}. The result is that $H_{mat}$ is of the form 

\begin{equation}
H_{mat} = \int dx \frac{-e_t^a\eta_{ab}e_x^b}{\sqrt{|e_x\cdot e_x|}}\mathcal{P}_0 + \int dx \frac{|e_t^a\epsilon_{ab}e_x^b|}{\sqrt{|e_x\cdot e_x|}}\mathcal{H}_0 
\label{matterhamiltonian}
\end{equation}

\noindent where

\begin{equation}
\mathcal{P}_0 = - \frac{\pi \partial_x \phi}{\sqrt{|e_x\cdot e_x|}} 
\end{equation}

\noindent and the function $\mathcal{H}_0$ is of the form

\begin{equation}
\mathcal{H}_0 = \sqrt{|e_x\cdot e_x|}\tilde{f}\left(\frac{\pi}{\sqrt{|e_x\cdot e_x|}},\frac{\partial_x \phi}{\sqrt{|e_x\cdot e_x|}},\phi\right). 
\end{equation}

\noindent Here $\tilde{f}$ is some function whose details depend on the particular form of $K$ (see Appendix \ref{appscalarham} for more details) and whose only importance for us here is to specify the dependence of $\mathcal{H}_0$ on the canonical variables.\\
\indent Crucially for what follows --as explained in more detail in Appendix \ref{appscalarham}-- it should be noted that 

\begin{equation}
\int dx \mathcal{H}_0 = E_0(L)
\label{undeformedenergyL}
\end{equation}

\noindent is the energy of the undeformed theory in flat space when placed on a spatial circle of size

\begin{equation}
L = \int dx \sqrt{|e_x\cdot e_x|}.
\label{dynamicallength}
\end{equation}

\noindent Similarly

\begin{equation}
\int dx \mathcal{P}_0 = P_0(L) 
\label{undeformedmomentum}
\end{equation}

\noindent is the spatial momentum density of the undeformed scalar theory when placed on a spatial circle of size $L$.\\
\indent We therefore find, as expected for a diffeomorphism-invariant theory, that the Hamiltonian is a linear combination of the constraints: 

\begin{equation}
\label{hamiltonian}
H = \int e_t^a F_a
\end{equation}  

\noindent where the $e_t^a$ play the role of Lagrange multipliers as anticipated and the $F_a = 0$ are the constraints of the theory, with\footnote{Here we have lifted the absolute value of $|e_t^a\epsilon_{ab}e_x^b|$ under the assumption that $e_t^a\epsilon_{ab}e_x^b>0$. This is merely a convention that somewhat simplifies what follows, as changing the sign of this quantity is equivalent to changing the sign of $\lambda$ and of the definition of $\mathcal{P}_0$.}

\begin{equation}
F_a \equiv \epsilon_{ab}\frac{\left(\partial_x X^b + e^b_x\right)}{\lambda}+ \frac{-\eta_{ab}e_x^b}{\sqrt{|e_x\cdot e_x|}}\mathcal{P}_0 +  \frac{\epsilon_{ab}e_x^b}{\sqrt{|e_x\cdot e_x|}}\mathcal{H}_0.
\label{constFa}
\end{equation}

To summarize our discussion thus far, from the canonical point of view the gravitational $T\bar{T}$ theory consists of a constrained Hamiltonian system. The unconstrained phase space, which we will denote $M_{kin}$, is described by the canonical variables $\{e_x,X,\phi,\pi\}$ and is imbued with the symplectic form $\Omega$ given by \eqref{symplecticform}. It is over this phase space that the constraints $F_a = 0$ are imposed.\\
\indent We now turn to the analysis of this constrained Hamiltonian system. The first important aspect of this is the fact that, as expected for a gravitational theory, the constraints are first class. Given the generality of the matter action we consider, it is worthwhile to discuss this in some detail, even if this is a completely expected result.\\ 
\indent It turns out that in order to analyze their structure, it is convenient to consider a redefined form of the constraints which we take to be

\begin{align}
    C_1 &\equiv -F_b\epsilon^b_ae^a_x=\left( e_x^a \eta_{ab} (e^b_x + \partial_x X^b) + \mathcal{H}_0 \right)\\
    C_2 &\equiv e^a_xF_b = \left( e^a_x \epsilon_{ab} \partial_x X^b + \mathcal{P}_0 \right).
\end{align}

\noindent Notice that, since we are assuming that the $e_x$, seen as a vector in the Minkowski-signature auxiliary space, is non-lightlike everywhere, the conditions $C_i=0$ are equivalent to the original constraints $F_a = 0$ (there is an invertible relation between the $C$s and $F$s under this assumption). Observe that if $e_x$ were lightlike at a given point, the induced metric on the spatial slice would be singular at that point. In fact, many of the previous formulas are only sensible within the assumption that this does not happen.\\
\indent The advantage of the formulation in terms of the $C_i$ over the $F_a$ is that, the $C_i$ being (at least) quadratic in the fields, their action via Poisson brackets on the canonical variables starts with a linear term instead of a constant (as happens for the $F_a$), and thus their Poisson brackets have a better chance of closing on a more manageable algebraic structure.\\
\indent Furthermore, in order to check their brackets it is more practical to define the smeared constraints

\begin{align}
    \mathcal{C}[N] &= \int dx N C_1 \label{constraintC}
    \\
    \mathcal{C}_x [N^x] &= \int dx N^xC_2 \label{constraintCx}.
\end{align}

\noindent defined for some arbitrary functions $N$ and $N_x$. Notice that this is equivalent to performing a redefinition of the Lagrange multipliers  
\begin{align}
    e^a_t \longrightarrow N^x e_x^a + N e_x^b \epsilon^a_b.
    \label{redefLagMult}
\end{align}
\noindent where $N$ and $N^x$ are taken to be independent from the canonical variables. With this redefinition, \eqref{hamiltonian} becomes:
\begin{align}
    H= \mathcal{C}[N] + \mathcal{C}_x [N^x].
    \label{hamSmeared}
\end{align}
\noindent Although this redefinition does not look too covariant, this is not an issue since in the canonical approach explicit covariance is broken by picking a particular foliation of spacetime.\\
\indent Let us examine the Poisson brackets between the smeared constraints. A simple calculation shows that the smeared constraint $\mathcal{C}_x [N^x]$ generates, via its Poisson bracket, spatial diffeomorphisms parametrized by the spatial vector field $N^x$. Therefore, all Poisson brackets involving $\mathcal{C}_x [N^x]$ are obtained trivially. The only bracket that is left is that of $\mathcal{C}[N]$ with itself, which is somewhat subtle due to the degree of generality we are assuming for the matter action. The details are presented in appendix \ref{appConstAlg}. The end result is:
\begin{align}
    \left\lbrace \mathcal{C}[N] , \mathcal{C}[M] \right\rbrace &= \mathcal{C}_x [N \partial_x M - M \partial_x N] 
    \label{PoissonCC}
    \end{align}
\begin{align}
    \left\lbrace \mathcal{C}_x [N^x] , \mathcal{C}[N] \right\rbrace &= \mathcal{C}[N^x \partial_x N - N\partial_x N^x] \label{PoissonCxC}
\end{align}
\begin{align}
\left\lbrace  \mathcal{C}_x [N^x] , \mathcal{C}_x [M^x] \right\rbrace &= \mathcal{C}_x [M^x \partial_x N^x - N^x \partial_x M^x] \label{PoissonCxCx}
\end{align}

\indent We see then that the $C_i$ constraints are indeed first class. Since the $F_a$ can be obtained from the $C_i$ by an invertible linear transformation then this implies that the $F_a$ are also first class.\footnote{In the terminology of constrained Hamiltonian systems, the $F_a$ would be secondary constraints. The primary constraints are those corresponding to the vanishing of the conjugate momenta to the $e^a_t$ variables. The $F_a$ are secondary constraints in that they result from the requirement that the primary constraints be conserved. The fact that these are first class and that the Hamiltonian is of the form \eqref{hamSmeared} ensures that no tertiary constraints arise. The distinction between primary, secondary and tertiary constraints is, however, irrelevant at the end of the day (see, e.g. \cite{henneaux}).}.\\
\indent As it usually happens for constrained Hamiltonian systems, first class constraints gen\-er\-ate gauge transformations on functions of the canonical variables via their Poisson brackets. In this case, the $\mathcal{C}_x[N^x]$ generate spatial diffeomorphisms and the $\mathcal{C}[N]$ generate the evolution along the time-like direction\footnote{Modulo a spatial diffeomorphism}.\\
\indent Regarding the constraint algebra obtained, a few final comments are in order. First, the constraint algebra obtained here is the usual one present in the canonical formulation of string theory and $1+1$ dimensional gravity theories (see for example \cite{Brink:1988nh} or \cite{jackiw1995two}). Sec\-ond, there are no structure functions present in the right side of the Poisson Brackets, only structure constants (so the algebra obtained here is thus a true Lie Algebra); the presence of structure functions in the constraint algebra is one of the main obstacles in the canonical quantization of gravity, since they imply that the group associated to the action of the constraints is non compact thus rendering the group averaging procedure to find physical states non-viable (See e.g. \cite{rovelli2004quantum,thiemann2008modern}). Finally, it is worth noting that the previous result was obtained considering matter consisting of a scalar field with a fairly general form of the action, having assumed only that the Lagrangian density depends on Lorentz invariant combinations of the scalar field and its derivatives up to second order. We haven't been able to find analyses of the constraint structure of gravitational systems with matter in the degree of generality considered here in the literature, and for this reason we present this calculation in detail in appendix \ref{appConstAlg}.\\
\indent We now turn to the description of the physical observables of the theory. From the point of view of the canonical formalism these will be what are known as Dirac observables: functions of the phase space variables that commute with the constraints of the theory. In general, these will to a certain degree depend on the particularities of the matter sector. However, in the case of $T\bar{T}$ gravity there are observables associated to the symmetries of the target space (which is the ``physical'' space of the theory) that are present independently of the nature of the matter sector. Indeed, examining \eqref{ttbargravityaction} it is clear that the theory is symmetric under the action on $X$ of the isometries of the target-space metric. In the case of constant $E_a$ which we are restricting to here, these will be Poincarè transformations of target-space. The conserved quantities associated to these symmetries are given by\footnote{Notice that the $X$ equations of motion imply $de^a=0$, and so $\int_{\mathcal{C}}e^a$ depends only topologically on the curve $\mathcal{C}$.}

\begin{align}
P^{a} &\equiv \int dx \frac{\epsilon^a_be^b}{\lambda}\label{targetspacemomentum}\\
J &\equiv  -\int dx\frac{Y^ae^b_x\eta_{ab}}{\lambda}
\end{align}

It is easy to check that the Poisson algebra of these quantities represents the Poincar\'e algebra

\begin{align}
\left\{J,P^{a}\right\}&=\epsilon^a_bP^b\\
\left\{P^{a},P^{b}\right\}&=0
\end{align}

\noindent and that these generate Poincaré transformations of the $X^{\mu}$ through the Poisson brackets

\begin{align}
\left\{X^{a},J\right\}&= \epsilon^a_bX^b\\
\left\{X^{a},P^{b}\right\}&=\eta^{ab}
\end{align} 

\noindent On the $e^a$ these act as

\begin{align}
\left\{e^{a}_x,J\right\}=\epsilon^a_be^a_x\\
\left\{e^{a}_x,P^{a}\right\}=0.
\end{align}

\noindent and they Poisson commute with the matter fields.\\ 
\indent From the canonical action of the $P^{a}$ on the fields it is clear that they are indeed Dirac observables of the theory, as they generate constant shifts of $X^a$, which appear with derivatives in the constraints.  On the other hand, $J$ is only a Dirac observable when periodic boundary conditions are taken for the $X^a$, but not in the winding sector, as expected, given that compactifying the target-space spatial slice breaks target-space boosts.\\
\indent An important observation for what follows is that $P^0$ and $P^1$ generate target-space translations and are thus the target-space energy $E$ and momentum $P$ respectively, which we crucially interpret as the $T\bar{T}$-deformed energy and momentum. The key point that we will show is that the constraints of the gravitational theory imply nontrivial relations between these observables that are equivalent to the finite volume $T\bar{T}$ flow equation \eqref{burgers}.\\
\indent In order to do this, it will be useful to perform a gauge-fixing procedure, that will allow to carve out the reduced phase space (in a sense, the ``physical'' phase space) out of the full phase space of the theory $M_{kin}$. As we have recalled earlier, in a Hamiltonian system with first-class constraints these are interpreted to generate via their Poisson action infinitesimal gauge transformations of the system. Thus to remove ambiguities coming from gauge evuivalences, the reduced phase space of the system is obtained by quotienting the full phase space by the action of these gauge transformations. One way to accomplish this is to supplement the constraints of the theory with a set of extra constraints --the gauge-fixing conditions-- so that imposing the full set of constraints picks out a single representative of each gauge orbit. Notice that the specific way these gauge-fixing conditions are imposed does not affect Dirac observables since they --by definition-- Poisson-commute with the first class constraints and are therefore gauge invariant. Indeed we will see that the $T\bar{T}$ finite volume flow equations arise from the constraints as gauge-invariant algebraic relations between gauge-invariant observables and thus are independent of the particular gauge-fixing procedure will follow.\\
\indent We will perform the gauge-fixing procedure in two steps and start by imposing first the partial gauge-fixing condition

\begin{equation}
\partial_xe^a_x = 0
\label{partialgaugefix}
\end{equation} 

\noindent Notice that indeed this does not fully fix the gauge freedom of the system given that the zero modes of the constraints Poisson-commute with this condition

\begin{equation}
\left\{\int dx F_a , \partial_xe^b_x \right\}=0.
\end{equation}  

\noindent However, condition \eqref{partialgaugefix} does fully fix the gauge transformations generated by the non-zero modes. To see this, let us define the expansion of the constraints and gauge-fixing conditions over the non-zero modes

\begin{align}
G^a_{(n)}\equiv \int dx G^a(x)e^{ix2\pi n}\hspace{20pt}n\neq 0\\
F_{a(m)}\equiv \int dx F_a(x)e^{ix2\pi m}\hspace{20pt}m\neq 0 
\end{align}
Then the matrix $G^b_{amn}$ of Poisson brackets between $G^a_{(n)}$ and $F^a_{(m)}$ is given by 

\begin{equation}
G^b_{amn}\equiv\left\{F_{a(m)} , G^b_{(n)}  \right\}=-\lambda m^2 \delta^a_b\delta_{mn}.
\end{equation} 

\noindent Let us now define the space $\tilde{M}_{kin}$ as that obtained from imposing the conditions $F^a_{(n)} = 0$ and $G^b_{(m)} = 0$ on the phase space $M_{kin}$. Then the fact that $G^b_{amn}$ is invertible implies that condition \eqref{partialgaugefix} does fix the gauge transformations generated by $F^a_{(m)}$ and that the closed differential form $\tilde{\Omega}$ induced on $\tilde{M}_{kin}$ is invertible and thus symplectic. Indeed, this symplectic form is given by  

\begin{equation}
\tilde{\Omega} = - \frac{1}{\lambda}\int dx \epsilon_{ab}\delta\tilde{X}^a \wedge \delta \tilde{e}^b_x + \int dx \delta\phi\wedge\delta \pi
\end{equation}

\noindent where the $\tilde{X}^a$ and $ \tilde{e}^b_x$ denote the zero modes for the corresponding variables. The partial gauge fixing procedure finally leaves us then with the phase space $\tilde{M}_{kin}$ on which we have two remaining constraints

\begin{equation}
\tilde{F}_a\equiv \int dx F_a \vert_{kin}=0
\end{equation}

\noindent where the $\vert_{kin}$ implies that this phase space function is to be evaluated at $F_{a(m)} = G^b_{(n)}=0$. \\
\indent We will now see how these left-over constraints imply non-trivial relations between the Dirac observables of the theory $P^a$ that are equivalent to the $T\bar{T}$ flow equation for the finite-volume spectrum.\\
\indent Notice first that on $\tilde{M}_{kin}$ we have 

\begin{equation}
    e_x^a\vert_{kin} = \tilde{e}_x^a = \lambda \epsilon^a_b P^b.
    \label{consteandP}
\end{equation}

\noindent If we now consider, as explained in the introduction, winding boundary conditions for $X^a$ where

\begin{equation}
    \int dx \partial_x X^a =  R\delta^a_1
\end{equation}

\noindent then the $\tilde{F}_a=0$ constraints can be written as 

\begin{equation}
\epsilon_{a1}\frac{R}{\lambda} + P_b + \sign(\lambda) \frac{P_a}{|P^2|^{1/2}}E_0(L) -\sign{\lambda}\frac{\epsilon_{ab}P^b}{|P^2|^{1/2}}P_0(L)=0
\label{tildeF}
\end{equation}

\noindent with $L$, $E_0(L)$ and $P_0(L)$ defined as in \eqref{undeformedenergyL},\eqref{dynamicallength} and \eqref{undeformedmomentum}. We now identify $P^a$ with the $T\bar{T}$ deformed energy $E_{\lambda}$ and the momentum $P_{\lambda}$ as follows

\begin{align}
    P^0 = -R/|\lambda| +\sign(\lambda) E_{\lambda}(R)\\
    P^1 = -\sign(\lambda) P_{\lambda}(R)
\end{align}

\noindent and define $\theta_0$ so that

\begin{align}
    \cosh \theta_0 = -\frac{P^0}{|P^2|^{1/2}} = \frac{R - \lambda E_{\lambda}(R)}{L}\\
    \sinh \theta_0 = -\frac{P^1}{|P^2|^{1/2}} = \lambda \frac{P_{\lambda}(R)}{L}
\end{align}

\noindent where combining \eqref{dynamicallength} with \eqref{consteandP} we have

\begin{equation}
    L = |\lambda||P^2|^{1/2} =  |(R-\lambda E_{\lambda}(R))^2 - P_{\lambda}(R)^2|^{1/2} 
\end{equation}

\noindent (notice that the ``physical'' sign for $\lambda$ in our conventions is negative). Then the $\tilde{F}_a=0$ equations can be written as

\begin{align}
    E_{\lambda}(R) &= \cosh\theta_0 E_0(L) - \sinh\theta_0 P_0(L)\label{implicitTTflowE}\\
    P_{\lambda}(R) &= -\sinh\theta_0 E_0(L) + \cosh\theta_0 P_0(L)\label{implicitTTflowP}.
\end{align}

These are precisely the equations that define the implicit form of the solution to the $T\bar{T}$ finite-volume energy flow equation \eqref{burgers}, as can be found in \cite{Cavaglia:2016oda,Conti:2018jho}. These is one of the main results of this paper: that the constraints arising from the gravitational system imply relations between the target-space energy and momentum that are equivalent to the $T\bar{T}$-deformed spectrum. Even though it was necessary to fix (partially) the gauge to arrive at these equations, these are algebraic relations between gauge-invariant observables and are thus gauge-invariant.\\
\indent Following \cite{Cavaglia:2016oda,Conti:2018jho}, let us further analyze the flow equations \eqref{implicitTTflowE} and \eqref{implicitTTflowP} to clarify their meaning. Noticing that they are of the form of a Lorentz transformation of rapidity $\theta_0$ between $(E_0(L),P_0(L))$ and $(E_{\lambda}(R),P_{\lambda}(R))$, it is straightforward to invert these equations into

\begin{align}
    E_0(L) &= \cosh\theta_0 E_{\lambda}(R) + \sinh\theta_0 P_{\lambda}(R)\\
    P_0(L) &= \sinh\theta_0 E_{\lambda}(R) + \cosh\theta_0 P_{\lambda}(R).
\end{align}

\noindent The last of these equations can be written as

\begin{equation}
    L P_0(L) = R P_{\lambda}(R) \label{undefMoment}.
\end{equation}

\noindent Since in finite spatial volume spatial momentum is quantized, then for any given state there are integers $N_0$ and $N_{\lambda}$ such that $P_0(L) = 2\pi N_0/L$ and $P_{\lambda}(R) = 2\pi N_{\lambda}/R$. Equation \eqref{undefMoment} then immediately implies that $N_0 = N_{\lambda}$, so that the integer describing the quanta of total spatial momentum is undeformed by the flow, as expected for an integer-quantized quantity in the presence of a continuous deformation. In the zero momentum sector, the deformation takes a particularly simple form

\begin{equation}
    E_{\lambda}(R) =  E_0(R -\lambda E_{\lambda}(R))
\end{equation}

\noindent which is the well-known implicit solution of \eqref{burgers} when $P=0$.\\
\indent It is interesting to note that the $T\bar{T}$ flow equations between $E_{\lambda}(R)$ and $E_0(R)$, which are supposed to be exact quantum mechanically, already arise at the classical level within the gravitational description we have presented. This hints at the fact that perhaps quantum effects in this gravitational theory are such that to take them into account one needs only account for them at the classical theory. That is, that to account for quantum corrections to $E_{\lambda}$ one merely needs to include quantum effects in the calculation of $E_0$, and then the quantum mechanical $E_{\lambda}$ emerges from taking this (quantum) $E_0$ as input into the classical flow equations. We discuss this point a bit further in section \ref{quantum}.\\
\indent To finish our analysis of the canonical structure of the theory we will construct the fully gauged-fixed ``reduced'' phase-space. As discussed above, the gauge-fixing condition \eqref{partialgaugefix} does not fully fix the gauge: transformations infinitesimally generated by the zero modes of the constraints remained unfixed. To obtain the fully gauge-fixed Hilbert space we have to supplement \eqref{partialgaugefix} with extra conditions that remove this leftover gauge freedom. we will now briefly discuss two possible ways in which this can be achieved.\\
\indent One possibility is to fix the gauge in order to simplify the on-shell behavior of the metric. The $X^a$ equations of motion in the spacetime split we have chosen are

\begin{equation}
    \partial_t e^a_x  - \partial_x e^a_t = 0.
\end{equation}

\noindent Combined with the gauge-fixing condition \eqref{partialgaugefix} these EOMs imply that $\partial_x e^a_t$ is only a function of $t$ only, meaning that $e^a_t$ is of the form

\begin{equation}
    e^a_t = x \partial_x e^a_t (t) + g^a(t)
\end{equation}

\noindent for some functions $g^a(t)$. If we impose that $e^a_t$ is periodic on the spatial circle then this periodicity requires the first term in the previous expression to vanish, and thus $e^a_t$ is independent of $x$ and only a function of $t$ in the gauge we have chosen. The remaining gauge freedom is generated by two functions (the zero modes of the $F_a$) and thus can be used to fully fix to constants the two functions $g^a(t)$ that determine $e^a_t$ on-shell. This is another way of seeing that $e^a_{\mu}$ can be gauge-fixed to be on-shell flat (indeed constant on the cylinder), which was one of the key points used in \cite{dubovsky2018tt} to calculate the quantum partition function and spectrum of this theory. Notice that the matter equations of motion in this gauge are the undeformed equations of motion in the presence of a constant metric on the cylinder, and thus the quantities $E_0(L)$ and $P_0(L)$ that appear in the $T\bar{T}$ flow equations are obviously constant in this gauge.\\
\indent From the canonical point of view, the gauge-fixing outlined in the previous paragraph is somewhat counter intuitive, as one is gauge-fixing Lagrange multipliers and not canonical variables, and so the fully-constrained phase space and induced symplectic structure are not easy to construct from this. Another perhaps more natural possibility from the canonical point of view is to take the gauge-fixing condition

\begin{equation}
    \tilde{X}^a = 0
    \label{zeromodegaugefix}
\end{equation}

\noindent From the point of view of the matter equations of motion, this condition has the disadvantage that these will involve a time dependent metric. However, in this gauge the structure of the fully-fixed phase space $M_{red}$ is far more evident. Indeed, in this gauge $M_{red}$ is identical to the phase space of the unconstrained matter theory --in this case a simple real scalar--  and the induced symplectic form is just the symplectic form of the matter sector. The gravitational sector is completely removed by the gauge fixing conditions and equations of motion.

\section{Free massless scalars and non-critical strings} \label{sec3massless}

As a particular example of the procedure we have discussed, it is instructive to consider the simplest possible case, namely, that of a system of $N$ free massless scalars. This is particularly interesting because, as it is well-known, the $T\bar{T}$ deformation of free massless scalars appears to be related to non-critical strings. A number of previous studies have discussed this (see \cite{Smirnov:2016lqw,Cavaglia:2016oda,Dubovsky:2012wk,Tolley:2019nmm,Callebaut:2019omt,Frolov:2019nrr} for a sample of various perspectives on this) and we will discuss this mostly to give a simple concrete example of the ideas discussed in the previous section.\\ 
\indent The matter sector of the action we consider is therefore simply\footnote{In our normalization whatever factor in front of the matter action may be reabsorbed into $\lambda$ without altering the classical equations of motion.}
\begin{align}
    S_0 = \int \mathrm{d}t \mathrm{d}x \sqrt{-g} g^{\mu\nu} \partial_\mu \phi^I \partial_\nu \phi^I
\end{align}

\noindent where $I=1\dots N$. The matter hamiltonian and momentum densities are %Performing the Legendre transform, using $g_{\mu\nu}=e^a_\mu e^b_\nu \eta_{a b}$ and choosing the same form for the Lagrange multipliers as we did in section \ref{classical}, we find that the matter contribution to the Hamiltonian and diffeomorphism constraints are given, respectively, by:
\begin{align}
    \mathcal{H}_0 &= \frac{\left( \pi^I\pi^I + \partial_x \phi^I\partial_x \phi^I \right)}{2|e_x\cdot e_x|^{1/2}} = \frac{h_0}{|e_x\cdot e_x|^{1/2}}
    \\
    \mathcal{P}_0 &= \frac{\pi \partial_x \phi}{|e_x\cdot e_x|^{1/2}} = \frac{p_0}{|e_x\cdot e_x|^{1/2}}
\end{align}

\noindent (we have defined $h_0$ and $p_0$ for later convenience) so that the total Hamiltonian becomes:
\begin{align}
    H = \int \mathrm{d}x   \left\lbrace N \left[\frac{\left( e_x^a \eta_{ab} (e^b_x + \partial_x X^b) \right)}{\lambda} + \left( \frac{1}{2} \pi^2 + \frac{1}{2} \left( \partial_x \phi \right)^2  \right) \right] + N^x \left( \frac{e^a_x \epsilon_{ab} \partial_x X^b}{\lambda} + \pi \partial_x \phi \right) \right\rbrace  
    \label{totalHamString}
\end{align}

%For this action, one finds that, in addition to the already mentioned conserved quantities from section (?):
%\begin{align}
%    P^a &= \int \mathrm{d}x \frac{\epsilon^{a}_b e^b_x}{\lambda}
%    \\
%    J &= - \int \mathrm{d}x  \frac{Y^a \eta_{a b} e_x^b}{\lambda}
%    \\
%    \mathcal{J} &= \int \mathrm{d}x \frac{Y^a \epsilon_{ab} e_x^b}{\lambda}
%\end{align}
%there are now two additional charges:
It is interesting to note that the $ISO(1,1)$ algebra of target-space isometries described in the previous section is now extended to $ISO(1,N+1)$. This is as not surprising from the fact that at the classical level this theory is expected to describe a bosonic string propagating in flat $(N+2)$-dimensional spacetime, with the fields $\phi^I$ realizing the embedding into the extra $N$ coordinates. The extra generators allowing to go from $ISO(1,1)$ to $ISO(1,N+1)$ are given by

\begin{align}
\label{charge_mat_translation}
    P^I &= \int \mathrm{d}x ~ \pi^I
    \\
    \label{transrot}
    Q^{IJ} &= \int dx\left(\pi^I\phi^J-\pi^J\phi^I\right)\\
    \label{charge_mat_grav_mix}
Q^{aI} &= \int dx\left(\phi^I ~ \left(e_x^a + \frac{\partial_x X^a}{2} \right) - \pi^I \epsilon^a_b \frac{X^b}{2}\right) %QUIERO PONER LAS DOS EN UJNA SOLA DEFINICION PERO NO ENCONTRE COMO HACERLO CON EPSILON Y ETA
\end{align}
Then $P^I$ and $Q^{IJ}$ generates translations and rotations in the $\phi^I$ sector, whereas $Q^{aI}$ has the following action on the canonical variables
%%%%%%%%%%%%REVISAR%%%%%%%%%%%%%%%%%%
\begin{align}
    \left\lbrace X^a , Q^{bI} \right\rbrace &= \epsilon^{a b} \phi^I 
    \\
    \left\lbrace e_x^a , Q^{bI} \right\rbrace &= \eta^{a b} \frac{\pi^I}{2} - \epsilon^{a b} \left(\frac{\partial_x \phi^I}{2} \right)
    \\
    \left\lbrace \phi^{I} , Q^{aJ} \right\rbrace &= \delta^{IJ}\epsilon^a_b \frac{X^b}{2}
    \\
    \left\lbrace \pi^I , Q^{aJ} \right\rbrace &= -\delta^{IJ}\eta^a_b \left( e_x^b + \frac{1}{2} \partial_x X^b \right)
\end{align}
so that the action of $Q^{aI}$ is a rotation/boost between the $X^a$ variables and the $\phi^I$. A straighforward calculation shows that indeed the poisson brackets of the generators $\left\{P^a,P^I,Q,Q^{IJ},Q^{aI}\right\}$ close into the $ISO(1,N+1)$ algebra. This shows that indeed at the classical level the theory realizes the isometries of flat $N+2$-dimensional spacetime. We interpret this theory to give a kind of ``Polyakov-like'' formulation of a non-critical bosonic string, where $X^{\mu}=\left\{X^a,\phi^I\right\}$ give the embedding coordinates and $e^a_x$ plays the role of a ``Polyakov metric''. Interestingly, in contrast to what happens in the usual Polyakov formulation of the bosonic string, in this formulation the dynamical geometry on the worldsheet $e^a_{\mu}$ is also charged under target-space transformations.\\
%Calculating the constraints between the conserved charges, one obtains:
%\begin{align}
%    \left\lbrace Q^a, P^b \right\rbrace &= -\epsilon^{a b} \Pi & \left\lbrace P^a , \Pi \right\rbrace &= 0
%    \\
%    \left\lbrace Q^a, \Pi \right\rbrace &= \delta^a_b P^b & \left\lbrace P^a , J \right\rbrace &= \epsilon^a_b P^b
%    \\
%    \left\lbrace Q^a , J \right\rbrace &= \epsilon^a_b Q^b & \left\lbrace \Pi , J \right\rbrace &= 0
 %   \\
 %   \left\lbrace Q^a , Q^b \right\rbrace &= \epsilon^a_b J & \left\lbrace \Pi , \Pi \right\rbrace &= 0
 %   \\
 %   \left\lbrace P^a , P^b \right\rbrace &= 0 & \left\lbrace J , J \right\rbrace &= 0
%\end{align}
\indent Let us now analyze the spectrum of this theory according to the analysis outlined in the previous section. We will consider that we are as before in the winding sector in $X^1$ so that we have an $N+2$-dimensional closed string winding around a compact target-space dimension of size $R$, with $\phi^I$ being the transverse coordinates. from our previous arguments, we start from the constraint equations, which formulated in terms of \eqref{totalHamString} appear naturally in the $C_i=0$ form

\begin{gather}
    \lambda P^2 - P^a \epsilon_{ab}\partial_x X^b + h_0 = 0\\
    P^a\eta_{ab}\partial_x X^b - p_0 = 0
\end{gather}
    
\noindent where we have already formulated the constraints in the $\partial_xe^a_x=0$ gauge and identified the $P^a$ as in \eqref{consteandP}. The zero mode equations in the winding sector then yield

\begin{gather}
    \lambda P^2 - P^0 R + \int dx h_0 = 0\label{zeromodeconstfreepart1}\\
    P^1 = \frac{\int dx p_0}{R}
    \label{zeromodeconstfreepart2}
\end{gather}

\noindent which combine to give $P^0$ of the form

\begin{equation}
    P^0 = \frac{R}{2\lambda} \pm \frac{1}{2\lambda}\sqrt{R^2 - 4\lambda \left(\int dx h_0 -\lambda \left(\frac{\int dx p_0}{R}\right)^2\right)}
    \label{P0forscalars}
\end{equation}

\noindent which yields the target-space energy spectrum of the string in terms of the spectrum of the operators $\int dx h_0$ and $\int dx p_0$, which are the flat-space energy and momentum of a system of $N$ free two-dimensional massless bosons on a compactified spatial circle of size $1$. Notice that in our formulation from section \ref{classical} the constraint equations giving rise to the $T\bar{T}$ deformed spectrum were naturally written in terms of the $E_0(L)\equiv \int dx \mathcal{H}_0$ and $P_0(L)\equiv \int dx \mathcal{P}_0$. For the simple system we consider in this section the relation between $E_0(L)$ and $P_0(L)$ and $\int dxh_0$ and $\int dxp_0$ is trivial at constant $e_x$ and the equation for the spectrum is readily obtained in terms of the latter.\\
\indent The undeformed theory of free massless bosons is readily solved at the full quantum level and its spectra is expressed in terms of two non-negative integers $N_L$ and $N_R$ describing the total number of left- and right-moving momentum

\begin{align}
    \int dx h_0 &= 2\pi N_R + 2\pi N_L - 2\pi N/12 \\
    \int dx p_0 &= 2\pi N_R - 2\pi N_L.
\end{align}

\noindent Plugging this into expression \eqref{P0forscalars} and taking $\lambda=-l_s^2/2$

\begin{equation}
    P^{0} = -\frac{R}{l_s^2} \mp \frac{1}{l_s}\sqrt{\frac{R^2}{l_s^2} + 4\pi \left(N_L + N_R - N/12\right) + \frac{4\pi^2l_s^2(N_L-N_R)^2}{R^2}}.
\end{equation}

\noindent This is precisely the lightcone-quantized bosonic string spectrum in $D = N+2$ in the unit-winding sector (for the positive energy branch, and up to subtraction of some linear in $R$ terms). Notice that we have obtained this by plugging in the full quantum result for the undeformed free theory describing the transverse modes into the ``classical'' $T\bar{T}$ flow equations. As mentioned before, the fact that the $T\bar{T}$ deformation of $N$ free massless bosons yields this spectrum is well known and can be seen directly by solving Burgers' equation \eqref{burgers} with free theory spectra as initial condition. We have merely presented this in order to exemplify our more general procedure.\\
\indent Notice that even from our point of view it is hardly surprising that we have obtained the light-cone string spectrum, given that the zero modes of our gauge-fixed constraints, equations \eqref{zeromodeconstfreepart1} and \eqref{zeromodeconstfreepart2}, coincide with the zero modes of the Virasoro constraints

\begin{gather}
   \int dx \left( \partial_x X^{\mu}\partial_x X_{\mu} + \partial_t X^{\mu}\partial_t X_{\mu}\right) = 0\\
    \int dx\partial_t X^{\mu}\partial_x X_{\mu} = 0
\end{gather}

\noindent in lightcone gauge where in the winding sector one has\footnote{Here we are using the somewhat unconventional notation of $t$ and $x$ for the worldsheet coordinates.}

\begin{align}
    X^0 &= \left(l_s^2 P^0 + R\right)t + \displaystyle\sum_{n\neq 0}\left(\alpha^{-}_n e^{i n(x-t)} + \tilde{\alpha}^{-}_n e^{i n(x+t)} \right)\\
    X^1 &= l_s^2 P^1t + R x - \displaystyle\sum_{n\neq 0}\left(\alpha^{-}_n e^{i n(x-t)} + \tilde{\alpha}^{-}_n e^{i n(x+t)} \right) 
\end{align}

\noindent where we have introduced the slightly unconventional parametrization of target-space momentum in order to reproduce equation \eqref{zeromodeconstfreepart1} on the nose.\\
\indent A very important point to mention before moving on is that from our previous construction one may be misled into thinking that the spectrum we have obtained is that of a well-defined quantum bosonic string theory propagating in $N+2$ dimensions. Of course this is not the case, and it is well-known that this spectrum is incompatible with target-space Poincar\'e invariance except for $N = 24$ or $N=1$. In lightcone quantization, the Poincar\'e algebra is not realized quantum mechanically outside those dimensions, and indeed our comments about $ISO(1,N+1)$ generators were only about classical Poisson brakets and are not expected to carry on into the full quantum theory.

\section{$J\bar{T}$-type deformations} \label{JTsection}
The type of classical flow equations for the spectra obtained in the previous sections for the $T\bar{T}$ deformation from its gravitational description can be extended to $J\bar{T}$-type deformations.  These are deformations of theories with global $U(1)$ symmetries, and the deformations are built out of Lorentz-breaking bilinears built out of the $U(1)$-current and the stress-energy tensor. Deformations of this type have been the subject of a several studies (see for instance \cite{Guica:2017lia,Bzowski:2018pcy,Chakraborty:2018vja,Aharony:2018ics,Apolo:2018qpq,Bhattacharyya:2023gvg} for a small sample of these). As explained in \cite{Aguilera-Damia:2019tpe}, these deformations can be described in a ``gravitational'' manner analogous to that of the $T\bar{T}$ deformation by coupling the undeformed theories $S_0$ not only to the dynamical geometry $e^a_{\mu}$ but also to a dynamical $U(1)$ gauge field $A_{\mu}$. Again we consider a set-up where the undeformed theory consists of a scalar field, in this case a complex scalar charged under the aforementioned $U(1)$ symmetry. The action of the deformed theory is then of the form

\begin{equation}
    S^{J\bar{T}}=S_{grav}^{J\bar{T}}+S_{0}[D_{\mu}\bar{\phi}D_{\nu}\phi g^{\mu \nu},\bar{\phi}\phi],
\end{equation}

\noindent where the coupling of the undeformed theory to the gravitational sector is as before and the coupling to the $U(1)$ gauge fields is through the covariant derivatives
\begin{align}
    D_{\mu}\phi &\equiv \partial_{\mu} \phi + iA_{\mu} \phi \\
    D_{\mu}\bar{\phi} &\equiv \partial_{\mu} \bar{\phi} - iA_{\mu} \bar{\phi}.
\end{align}

Similarly to what happens with the $T\bar{T}$ case, the deformation is implemented by the ``gravitational'' action governing the dynamics of the $e$ and $A$ fields which we take to be form

\begin{equation}
    S_{grav}^{J\bar{T}} = S_{grav}^{T\bar{T}}+\frac{1}{\lambda}\int\partial_{\mu}X^a\epsilon_{ab}n^b\left(\partial_{\nu}\alpha - A_{\nu} \right) \epsilon^{\mu\nu}+ \frac{1}{\lambda} \int e_{\mu}^a\epsilon_{ab}n'^b\left(\partial_{\nu}\alpha - A_{\nu}\right)\epsilon^{\mu\nu}
    \label{SgravJT}
\end{equation}

\noindent where $\alpha$ is an auxiliary scalar playing a similar role to the $X^a$ in this framework and $n^a$ and $n'^a$ are fixed vectors in target-space. Notice that this action breaks target-space Lorentz invariance, as expected for these types of Lorentz-breaking deformations. Notice also that in \cite{Aguilera-Damia:2019tpe} the deformation involves two $U(1)$ gauge fields, as the authors consider the undeformed theory to be a CFT and the possibility of coupling separately to the right-moving and left-moving currents. Here we consider a somewhat different situation where the undeformed theory is an arbitrary theory with a $U(1)$ symmetry and attempt to build a Lorentz-breaking deformation out of a procedure inspired by the ``gravitational'' action suggested in \cite{Aguilera-Damia:2019tpe} but with only one gauge field.\\
\indent Taking this formulation as a starting point, we may proceed as in section \ref{classical} and extract from the constraints flow equations for the finite-volume observables of the theory. To the observables available before, associated to the generators of target-space translations, now is added a generator of the symmetry 

\begin{equation}
\alpha\longrightarrow \alpha + \text{const},  
\label{shiftalpha}
\end{equation}
 
\noindent which we interpret as the deformed $U(1)$ charge. We proceed in an analogous way to section \ref{classical} and sketch the relevant steps, commenting on where they differ to the earlier procedure. \\
\indent The symplectic form of the theory is extracted from the part of the action that is linear in time derivatives

\begin{equation}
    S_{t}^{J\bar{T}} = \frac{1}{\lambda} \int \epsilon_{ab}X^a \partial_{t}e_{x}^b +\frac{1}{\lambda} \int \partial_t X^a \epsilon_{ab} n^b A_X + \frac{1}{\lambda} \int e_x^a \epsilon_{ab} n'^b \partial_t \alpha.
\end{equation}

\noindent It is useful to bring this expression to Darboux form by the redefinition

\begin{equation}
    \tilde{X}^a := X^a+n'^a \alpha,
\end{equation}

\noindent and 

\begin{equation}
    \tilde{e}^a := e^b+n^b A_X.
\end{equation}

\noindent in terms of which the symplectic form becomes\footnote{We introduce the notation $n'\cdot\epsilon\cdot n \equiv n'^a\epsilon_{ab}\dot n^b$}

\begin{equation}
    \Omega^{J\bar{T}} = -\frac{1}{\lambda} \int \delta \tilde{X}^a \wedge \tilde{e}_{X}^b \epsilon_{ab} - \frac{1}{\lambda} \int \delta \alpha \wedge \delta A_X (n'\cdot\epsilon\cdot n), \label{symplecticJT}
\end{equation}

\noindent The Poisson brackets are then the following

\begin{align}
    \lbrace \tilde{X}^a,  \tilde{e}_X^b\rbrace&=- \lambda \epsilon^{ab}\\
    \lbrace \alpha,  A_X \rbrace &= - \frac{\lambda}{n'\cdot\epsilon\cdot n}
\end{align}
We can express the Hamiltonian of the full deformed theory in the same way we did in section \ref{classical}

\begin{equation}
    H^{J\bar{T}}=H_{grav}^{J\bar{T}}+H_{mat}^{J\bar{T}}
    \label{HamiltonianJT}
\end{equation}
with

\begin{align}
    H_{mat}^{J\bar{T}}&=\sqrt{-g} \tilde{F} \left(\frac{\pi_{\phi}}{\sqrt{-gg^{00}}},\frac{\pi_{\bar{\phi}}}{\sqrt{-gg^{00}}}, \frac{D_x\bar{\phi}D_x\phi}{\sqrt{-gg^{00}}}\right)+ A_t\left( i \bar{\phi} \pi_{\bar{\phi}} - i \phi \pi_{\phi} \right) \\ & - \frac{g^{01}}{g^{00}} (\partial_X \phi +i A_x \phi) \pi_{\phi} - \frac{g^{01}}{g^{00}} (\partial_X \bar{\phi} - i A_x \bar{\phi}) \pi_{\bar{\phi}} \label{HO_JT}
\end{align}

\noindent being the matter contribution, which is derived through the Legendre transform of the action using the standard approach.\\
\indent From \eqref{SgravJT} we can see that the gravitational part of \eqref{HamiltonianJT} is given by

\begin{equation}
    H_{grav}^{J\bar{T}}=H_{grav}^{T\bar{T}}+A_t \left ( \frac{\partial_X X^a}{\lambda} \epsilon_{ab} n^b + \frac{e_X^a \epsilon_{ab} n'^b}{\lambda}\right)
    + e_t^a \frac{\epsilon_{ab}n'^b}{\lambda} \left( \partial_X \alpha - A_X \right).
\end{equation}

\noindent (superindices $T\bar{T}$ indicate terms that are identical to the pure $T\bar{T}$ case). Again, the total Hamiltonian is a linear combination of constraints of the form
\begin{align}
    H^{J\bar{T}}= e_t^a F^{J\bar{T}}_a + A_t G^{J\bar{T}}
\end{align}
where

\begin{equation}
    G \equiv \left( \frac{\partial_x X^a}{\lambda} \epsilon_{ab} n^b +  \frac{e_x^a \epsilon_{ab} n'^b}{\lambda}  \right) -  i (\pi_{\phi} \phi - \pi_{\bar{\phi}} \bar{\phi} )
\end{equation}
and
\begin{equation}
F_a \equiv \frac{\epsilon_{ab}}{\lambda}\left(\partial_x X^b + e^b_x + n'^b[\partial_x \alpha - A_X]\right) - \frac{\eta_{ab}e_x^b}{\sqrt{|e_x\cdot e_x|}}\mathcal{P}_0^{J\bar{T}} -  \frac{\epsilon_{ab}e_x^b}{\sqrt{|e_x\cdot e_x|}}\mathcal{H}_0^{J\bar{T}}.
\label{constFaJT}
\end{equation}

Note that  $\mathcal{P}_0$ and $\mathcal{H}_0$  have the same form as in section \ref{classical}, but they are evaluated in the background of the gauge field $A_X$, so

\begin{equation}
\mathcal{P}_0^{J\bar{T}} = - \frac{\pi D_x \phi}{\sqrt{|e_x\cdot e_x|}} 
\end{equation}
and
\begin{equation}
\mathcal{H}_0^{J\bar{T}} = \sqrt{|e_x\cdot e_x|}\tilde{f}\left(\frac{\pi_{\phi}\pi_{\bar{\phi}}}{\sqrt{|e_x\cdot e_x|}},\frac{D_x \phi D_x \bar{\phi}}{\sqrt{|e_x\cdot e_x|}},\bar\phi \phi\right). 
\end{equation}

To summarize, as for the case of the $T\bar{T}$ deformation theory, the ~$J\bar{T}$ theory is also a constrained Hamiltonian system. The imposition of the constraints $F_a=0$  and $G=0$ occurs over the phase space described by the canonical variables $\lbrace e_x, X, A, \alpha, \phi, \pi_{\phi}, \bar{\phi}, \pi_{\bar{\phi}}\rbrace$ equiped with the symplectic form \eqref{symplecticJT}. In an analogous manner to section \ref{classical}, it can be verified that constraints are indeed first class, as expected. %To illustrate this, we will initiate the analysis with the redefinition \eqref{redefLagMult}  of the Lagrange multipliers, which gives rise to the expression {\color{red}tengo la impresion que esto es completamente innecesario a los efectos de hallar el espectro y era solo un truco para calcular facilmente las constraints, es decir yo ni incluiria a los Cs}

%\begin{align}
%    H^{J\bar{T}}= \mathcal{C}^{J\bar{T}}[N] + \mathcal{C}_x^{J\bar{T}} [N^x]
%\end{align}
%with {\color{red} Ojito que estos no los terminé}
%\begin{align}
%    \mathcal{C}^{J\bar{T}}[N] &= \int N blabla%\int N \left( e_x^a \eta_{ab} (e^b_x + \partial_x X^b) + \mathcal{H}_0 \right) 
%    \label{constraintC_JT}
%    \\
%    \mathcal{C}_x^{J\bar{T}} [N^x] &= \int N^x ble ble%\left( e^a_x \epsilon_{ab} \partial_x X^b + \mathcal{P}_0 \right) 
%    \label{constraintCx_JT}
%\end{align}

Let us now identify the Dirac observables of the theory. As previously mentioned, the theory is symmetric under the action of isometries of the target-space metric on $X$ and the shift \eqref{shiftalpha}. The conserved quantities associated with these symmetries are
 
\begin{equation}
    P^a = \int dx \frac{ \epsilon^a_b e^b_x - \epsilon^a_b n^b A_x}{\lambda} 
\end{equation}

\noindent and

\begin{equation}
    Q_{\lambda} = -\int dx \frac{n'\epsilon n}{\lambda} A_x,
\end{equation}

\noindent which are the generators of translations in $X^a$ and $\alpha$, respectively.

As before, we use diffeomorphisms and $U(1)$ gauge transformations to fix the gauge where $e^a_x$ and $A_x$ are constants (denoted by tildes), so 

\begin{align}
    \tilde{A}_x&= - \frac{\lambda Q_{\lambda}}{(n'\cdot \epsilon\cdot n)}\\
    \tilde{e}^a_x&=\epsilon^a_b P^b \lambda - \frac{\lambda Q_{\lambda}n^a}{(n'\cdot \epsilon\cdot n)}
\end{align}

Proceeding similarly to section \ref{classical}, we find that the zero modes of the  constraints lead to the following equations, which represent a deformation mixing the $P^a$ with the deformed charge $Q_{\lambda}$

\begin{align}
    Q_{\lambda} &= Q_0 + P\cdot n' + \frac{Rn^1}{\lambda} \\
    \frac{R}{\lambda} + P^0 - Q_{\lambda} \frac{n^1-n'^1}{n'\cdot\epsilon\cdot n} &= \cosh\theta_0E_0(L,\mu) + \sinh\theta_0P_0(L,\mu)\\
     P^1 - Q_{\lambda} \frac{n^0-n'^0}{n'\cdot\epsilon\cdot n} &= \sinh\theta_0E_0(L,\mu) + \cosh\theta_0P_0(L,\mu)
\end{align}

\noindent where we have defined

\begin{gather}
    L = |\tilde{e}_x^2|^{1/2} = |\lambda|\left\lvert\left(P^0-Q_{\lambda}\frac{n^1}{n'\cdot \epsilon\cdot n}\right)^2 - \left(P^1-Q_{\lambda}\frac{n^0}{n'\cdot \epsilon\cdot n}\right)^2 \right\rvert^{1/2}\\
    \cosh \theta_0 =\frac{\tilde{e}^1_x}{L} = \left(P^0-Q_{\lambda}\frac{n^1}{n'\cdot \epsilon\cdot n}\right)\frac{\lambda}{L}\\
    \sinh \theta_0 = \frac{\tilde{e}^0_x}{L} = \left(P^1-Q_{\lambda}\frac{n^0}{n'\cdot \epsilon\cdot n}\right)\frac{\lambda}{L}.
\end{gather}

\noindent Also, in the previous equations $Q_0$ stands for the $U(1)$ charge generator in the undeformed theory 

\begin{equation}
    Q_0 = i\int dx\left(\pi_{\phi}\phi - \pi_{\bar{\phi}}\bar{\phi}\right) 
\end{equation}

and 

\begin{align}
    E_0(L,\mu) = \int dx\mathcal{H}_0^{J\bar{T}}\big\rvert_{e_x=\tilde{e}_x} \\
    P_0(L,\mu) = \int dx \mathcal{P}_0^{J\bar{T}}\big\rvert_{e_x=\tilde{e}_x}
\end{align}

\noindent are the energy and momentum of the undeformed theory when placed on a spatial circle of length $L$ and in the presence of a background constant background $U(1)$ gauge field of the form

\begin{equation}
    A_{\alpha} =  \delta_{\alpha x}\mu = -\delta_{\alpha x}\lambda \frac{Q_{\lambda}}{n'\cdot\epsilon\cdot n}, 
    \label{backgf}
\end{equation}

\noindent which can be equivalently thought as a $U(1)$ twist of the boundary conditions for the scalars along the circle

\begin{equation}
    \phi(t, x+L) = e^{i\mu}\phi(t,x) 
\end{equation}

\noindent with $\mu$ defined in \eqref{backgf}.\\
\indent In principle it is not evident how to relate the flow equations we have obtained to those found for the $J\bar{T}$-deformations in the literature. This is because our expressions involve the undeformed charges evaluated at twisted boundary conditions, with the twist parameter itself depending on the deformed spectra, in an analogous way as happens for the radius in the usual $T\bar{T}$ flow. This feature --to our knowledge-- is absent from usual formulations of $J\bar{T}$-type flows.\\
\indent It is interesting to point our however that having the flow depend on the undeformed theory at twisted boundary conditions with twists depending on the deformed conserved quantities is precisely what was found in \cite{Hernandez-Chifflet:2019sua} for the flows induced by higher-spin currents in integrable theories. Even though these results were specifically about these types of deformations, it seems likely that this twist dependence is a general feature of deformations of integrable theories by CDD factors. In any case, we have presented this result mostly as an application of the ``gravitational'' approach to a different type of deformation and relegate further study of the flow equation we have obtained to future work. 

\section{From canonical quantization to the path-integral formulation}\label{quantum}
%{\color{red} no me gusta mucho este t\'itulo para esta secci\'on}\\
%\indent  We will now discuss the quantization of the theory. As with all constrained theories, there are a multitude of approaches towards quantizations of such systems, which can roughly be divided into two categories: the ``constrain first'' and the ``quantize first'' approaches.\\
%We will begin with the discussion of ``constrain first'' type of approach. The starting point of this approach is the fully gauge-fixed phase space $M_{phys}$ with its induced symplectic form. As discussed in the last few paragraphs of section \ref{classical}, in the gauge-fixing we have chosen the fully fixed phase space and induced symplectic form coincide with those of the undeformed theory {\color{red}no se si es una buena idea llevarlo a esta discusi\'on. tal vez lo importante sea discutir en paralelismo con las cuantizaciones de cuerdas}.\\
Let us see now how the covariant path integral quantization \eqref{torusPartfunCovariant} discussed in \cite{dubovsky2018tt} emerges from the canonical point of view. We will follow the standard procedure (as found in many textbook treatments of QED \cite{Itzykson:1980rh,weinberg1995quantum,henneaux}) which starts from canonical quantization of a fully constrained system and its canonical-variables path integral and builds from there the covariant path integral over the unconstrained configuration space.\\ 
\indent We will start from the assumption that we have managed to quantize the theory in a ``constrain first'' type of approach, meaning that we have built a fully constrained ``reduced'' or ``physical'' phase space $M_{phys}$ by imposing at the classical level all of the constraints of the theory and have performed the canonical quantization procedure of this phase space. To be concrete, we will assume that this phase space is constructed out of the gauge-fixing procedure discussed in the last paragraphs of section \ref{classical} by imposing the gauge fixing conditions \eqref{partialgaugefix} and \eqref{zeromodegaugefix}. As explained before, the phase space thus constructed is equal to the phase space of the undeformed theory imbued with the symplectic form of the undeformed theory. Therefore, we may well assume that a quantization of this phase space is available, that is, that the undeformed theory is defined at the quantum level. In particular the canonical quantization procedure applied to the undeformed theory would yield a Hilbert space $\mathcal{H}_{phys}$ and a way to promote phase space variables in $M_{phys}$ to operators (modulo ordering issues as usual) and our preceding argument implies that this quantizes the fully constrained ``gravitational'' theory as well.\\
\indent From the point of view of the constrained gravitational theory the $P^a$ are natural (commuting) observables realizing physical spacetime translations and it is natural then to write down the partition function\footnote{We are keeping the Minkowskian $i$ in the exponent whereas for the partition function euclidean signature would be more natural (as considered in \cite{dubovsky2018tt}). This is irrelevant for our discussion which is completely formal in any case.}

\begin{equation}
    Z(L,R) = \text{Tr}_{\mathcal{H}_{phys}}\left\{ e^{-i L_a \hat{P}^a}\right\}.
    \label{canonicalconstrpartitionfunction}
\end{equation}

\noindent This function is theoretically well-defined, given that on the constraint surface, $P^a$ can be written in terms of phase space variables of the undeformed theory for which we assume a canonical quantization procedure makes sense. The $\hat{P}^a$ denote the operators on $\mathcal{H}_{phys}$ that would be obtained from this procedure. The $R$ dependence appears because the equations $\int dxF_a = 0$ that are used to write $P^a$ in terms of the reduced phase space canonical variables depend on $R$ via the winding of $X$ along the spatial cycle, which we take to be non-zero.\\ 
\indent Standard manipulations allow to write the $Z(L,R)$ partition function as a path integral over the fully-constrained phase space 

\begin{equation}
Z(L,R) = \int_{z^i(t=1)=z^i(t=0)}Dz e^{i\int dt (\alpha^{(const)}_i \dot{z}^i - L_aP^a) }
\label{constphasespacepartfun}
\end{equation}

\noindent where the $z^i$ stand for the phase-space variables of the fully-constrained phase space in a Darboux coordinatization and $\alpha^{(const)}_i$ for the corresponding symplectic potential. Of course, as we have mentioned, these coincide with the undeformed phase-space variables and in our scalar field example thy would be represented by the $\left\{\phi,\pi\right\}$ canonical pair. It is important to point out that in the previous expression $P^a$ stand for phase space functions built out of the $\hat{P}^a$ operators according to some ordering procedure (for instance, following \cite{weinberg1995quantum}, by pushing all canonical momentum to the right). Notice that since the $P^a$ on the fully-constrained phase space are obtained by solving the constraint equations, these are very nontrivial functions over this phase space and ordering issues may be very complicated to resolve in practice. 
Keeping this in mind, we can follow the usual procedure (\cite{Itzykson:1980rh,henneaux}) and express this path integral as one over the unconstrained gravitational phase space by adding the suitable delta functions and measure factors

\begin{equation}
Z(L,R) = \int De_x^a DY^a D\phi D\pi \delta(F_a)\delta(G^b)\det\left\{F_a,G\right\}e^{i\int \alpha^{(unconst)} - i\int dt L_a P^a}
\label{partfununconstrphasespace}
\end{equation}

\noindent where all of the path-integral integration variables are subject to periodic boundary conditions along $t\in (0,1)$. Notice that our canonical variable is the periodic $Y$ and not $X$ as we are assuming we are in the sector with spatial winding so that $X = X^{wind}+Y$. Here $F_a$ are the constraints as previously discussed and $G^b=0$ stand for our gauge fixing conditions \eqref{partialgaugefix} and \eqref{zeromodegaugefix}. In the previous formula $\alpha^{(unconst)}$ stands for (the pullback to the circle of) the unconstrained symplectic potential which is given by

\begin{equation}
    \int \alpha^{(unconst)} = \frac{1}{\lambda}\int dt dx \epsilon_{ab}e^a_x\partial_t Y^{b} + \int dt dx \pi\partial_t\phi.  
\end{equation}

We can now write the delta function enforcing the constraints of the theory as an exponential by integrating-in a Lagrange multiplier which is immediately identified with  $e^a_t$

\begin{equation}
    \delta(F_a) = \int De^a_t e^{-i \int dtdx e^a_t F_a} = \int De^a_t e^{-i \int dt (H_{grav} + H_0)}. 
\end{equation}

\noindent The path integral therefore becomes 
\begin{align}
    Z(L,R) &=\nonumber\\ &\int DY^aDe_x^aDe_t^a \delta(G^b)\det\left\{F_a,G\right\}\times \nonumber\\
    &\exp\left\{\frac{i}{\lambda}\int dt dx \epsilon_{ab}e^a_x\partial_t Y^{b} -\frac{i}{\lambda}\int dtdx e^a_t \epsilon_{ab}\left(\partial_x X^b + e^b_x\right)- i\int dt L_a P^a\right\}Z_0[e]
\end{align}
\noindent where $X$ includes the part realizing the spatial winding and we have used expression \eqref{Hgrav} for $H_{grav}$. Here we have identified the undeformed partition function in the presence of the gravitational field

\begin{equation}
    Z_0[e] \equiv \int D\phi D\pi e^{i\int dt dx \pi\partial_t\phi - i\int dt H_0}
\end{equation}

\noindent where the dependence on the dynamical geometry enters through $H_0$ as given in \eqref{matterhamiltonian}. Notice that we have been using the notations for a scalar field, although we expect these rather formal manipulations to be valid for some more general cases. Finally, we can use \eqref{targetspacemomentum} to write $P^a$ in the exponent in terms of the  unconstrained canonical variables 

\begin{equation}
    \exp\left\{-i\int dt L_aP^a\right\} = \exp\left\{-i\int dt L_a\int dx \frac{\epsilon^a_b e^b_x}{\lambda}\right\} =  \exp\left\{i\int dt\int dxe^a_x \epsilon_{ab}\frac{\partial_t(tL^a)}{\lambda}\right\},
\label{expP}
\end{equation}

\noindent given that the path integral is actually non-zero only in the support of the delta function $\delta(F_a)$ that enforces the constraints. Notice that this last point is actually somewhat subtle: the $P^a$ that appears in the constrained phase space path integral is really some ordered version of an operator whose expression in terms of the constrained canonical variables is expected to be complicated. The fact that after the ordering procedure this function of the constrained canonical variables can still be written as in \eqref{expP} is not at all evident. Under the assumption that this is so, if we redefine the field $X$ so that now it also includes winding along the time cycle 

\begin{equation}
    X^a \longrightarrow X^a + tL^a 
\end{equation}

\noindent we then recover (up to a constant term coming from total derivatives) the covariant path integral on the torus

\begin{equation}
    Z(L,R) = \int DX^aDe^a \delta(G^b)\det\left\{F_a,G\right\}e^{iS_{grav}[X,e]}Z_0[e].
    \label{covariantPI}
\end{equation}

\noindent This is precisely a gauge-fixed form of the covariant path integral on the torus considered in \cite{dubovsky2018tt}, where the $X$ windings around the two cycles of the torus encode the $L$ and $R$ dependence. Crucially, this path-integral expression is in fact independent of the gauge-fixing conditions used, see for instance \cite{Itzykson:1980rh} for arguments directly within the canonical approach\footnote{This can also be seen by the usual BRST arguments.}. Notice that usually (e.g. in actions that are second-order in time derivatives) to get from the canonical path integral to the covariant path integral it is necessary to integrate out some conjugate momenta, whereas here the phase space integral already yields a covariant path integral after exponentiation of the constraints (which is natural from the fact that actions that are first order in time derivatives --such as this one-- naturally weight paths in phase space, not in configuration space).\\
\indent It is interesting to compare the path integral expressions obtained here with that used in \cite{dubovsky2018tt}. In \eqref{covariantPI}, the integration measures obtained from the canonical procedure and implied in $De$ and $DX$ are in a sense trivial, in that they simply run over all phase space values of $X^a$ and $e_x^a$ and all real values of $e_t^a$ and there is no extra measure factor attached except what would come from the $\delta(G^b)\det\left\{F_a,G^b\right\}$ factor. From the canonical point of view this is natural, as the only requirement for our measure over the fully-constrained phase space is that it is invariant under canonical transformations. Indeed the path integral measure in \eqref{constphasespacepartfun} can be thought of, at every time slice, as the symplectic measure of the induced symplectic form on the gauge-fixed phase space.\\
\indent This is in contrast with the situation in the gauge-fixed path integral considered in \cite{dubovsky2018tt}, where even aside from the determinant factors coming from the Faddeev-Popov procedure the path integral measures over $e$ and $X$ are non-trivial and defined via diff-invariant inner products on the space of fields (as is usual in path integral formulations of string theory \cite{Polchinski:1985zf,DHoker:1988pdl}). Indeed, in the treatment of \cite{dubovsky2018tt} these measure factors were crucial to get the ``correct'' $T\bar{T}$ answer. Furthermore, the range of integration of the fields considered in the path integral formulation of \cite{dubovsky2018tt} ran only over explicitly invertible geometries by construction\footnote{One could argue however that since that path integral ended up being one-loop exact in a sense, the localization on the saddle point makes questions regarding the region of integration somewhat subtle.} in the canonical formulation the langrange multiplier variable $e_t$ runs over all real values, including values corresponding to non-invertible metrics for instance.\\ 
\indent This discrepancy of the covariant and canonical approaches in the range of integration for metric variables imposing constraints is often present in quantization of gravitational theories\footnote{See for instance \cite{Witten:2022xxp} for recent comments on this}. In fact, this problem is already present for the simplest possible diffeomorphism invariant action, that of the free relativistic massive particle. As discussed for instance in \cite{Hartle:1986yu,Cohen:1985sm} the worldline ``lapse'' function emerges in the canonical formalism as a Lagrange multiplier, whose range is naturally unrestricted. However, for the path integral to correctly yield the Klein-Gordon propagator, its range of integration must be restricted ultimately to positive values\footnote{It is also interesting to compare statement in \cite{Hartle:1986yu} and \cite{Cohen:1985sm} regarding to what extent locality fixes the path integral measure}.\\
\indent It is also interesting to compare the two path integral formulations for the $T\bar{T}$ gravity theory regarding the one-loop exactness property found in \cite{dubovsky2018tt}. There, after the path integral has localized into the space of constant metrics (constant dyads, strictly speaking) the remaing finite-dimensional integral is also found to be one-loop exact. From the point of view of that procedure this one-loop exactness was somewhat surprising, although crucial to obtain the correct $T\bar{T}$ deformed spectrum. As we have pointed out before, considering this perspective, the emergence of the equation for the $T\bar{T}$ spectrum at the classical level t is not surprising, given that at the level of the quantum partition function it arises from some saddle-point equation.\\
\indent In the context of the analysis presented in this paper, the one-loop exactness of the path integral and the resulting $T\bar{T}$ spectrum follow from our canonical analysis and  the expression for the canonical partition function we start with \eqref{canonicalconstrpartitionfunction}. It is instructive to see how this one-loop exactness comes about, and as a sanity check of this whole procedure, to rederive the $T\bar{T}$ spectrum from \eqref{covariantPI} working backwards. From this point of view, the step where one would find the localization to constant geometries would be when arriving at \eqref{partfununconstrphasespace}. In order to analyze this expression it is convenient to separate the constraints $F_a$ and the gauge fixing conditions $G^a$ into zero modes (denoted with tildes) and non-zero modes (denoted with primes). For the gauge-fixing conditions, the non-zero mode part $G'a$ is the $\partial_x e^a_x = 0$ constraint, while the zero-mode part $\tilde{G}^a$ is given by $\tilde{X}^a = 0$ (as in \eqref{zeromodegaugefix}). Notice that since the Poisson bracket between the zero mode of $F_a$ and the non-zero mode of $G$ vanish, then the determinant factor in  \eqref{partfununconstrphasespace} factors into

\begin{equation}
    \det\left\{F_a,G\right\} = \det\left\{\tilde{F}_a,\tilde{G}\right\}\det\left\{F'_{a},G'^b\right\}
\end{equation}

\noindent where the second factor is a functional determinant given by

\begin{equation}
    \det\left\{F'_{a},G'^b_{(n)}\right\} = \det(-\partial^2_x)^2.
\end{equation}

\noindent To perform the path integral in \eqref{partfununconstrphasespace} it is useful to factor the integration into zero and non-zero modes as

\begin{align}
    DY^a = d\tilde{Y}^aDY'^a \\
    De_x^a = d\tilde{e}^a_xDe'^a_x
\end{align}

\noindent where the primed variables again refer to the non-zero modes. Notice that the integral over the non-zero modes have no extra measure factor coming from a non-trivial path integral measure as in \cite{dubovsky2018tt}. The integrals over the nonzero modes of $Y^a$ and $e^a_x$ are immediately taken care of by the non-zero-mode part of the constraints and gauge-fixing conditions respectively. The fact that this sets to zero the non-zero mode part of $e^a_x$ means that these variables (that is, their non-zero mode parts) disappear from the exponent in \eqref{partfununconstrphasespace} since this exponent is linear in $e^a_x$ and depends on $Y$ via its product with $e^a_x$. Then these integrations yield only left-over functional determinants which are given by

\begin{align}
    \int DY'^a \delta(F'_{a}) = \int DY'^a \delta(\partial_x Y^a + \dots) = \det(\partial_x)^{-2} \\
    \int De'^a_x \delta(\partial_xe^a_x) = \det(\partial_x)^{-2}
\end{align}

\noindent These functional determinants can be turned into determinants of hermitian operators by using

\begin{equation}
    \det(Q) = \det(Q^{\dagger}Q)^{1/2}
    \label{hermitiantrick}
\end{equation}

\noindent If we assume, as mentioned before, that the path integral measures of integration for the $Y$ and $e_x$ are defined by the trivial inner-product for functions on the circle

\begin{equation}
    (f,g) \equiv \int dx f^* g
\end{equation}

\noindent (notice that, as mentioned before, there is no need to use a diffeomorphism invariant inner product on the circle to define a path integral over the canonical variables) then, using this inner-product to define the adjoint, the aforementioned manipulations of functional determinants \eqref{hermitiantrick} yield 

\begin{equation}
    \det(\partial_x) = \det(-\partial_x^2)^{1/2}.
\end{equation}

\noindent Thus, the functional determinants coming from the integration over the non-zero modes cancel with the functional determinant factor from $\det\left\{F_a,G\right\}$. The integral over the zero mode of $Y$ trivially sets it to zero because of the zero mode of the gauge fixing condition and we are left with a finite-dimensional integral over constant $e_x^a$ and a path integral over the canonical variables of the undeformed theory

\begin{equation}
    Z(L,R) = \int D\phi D\pi \int d\tilde{e}^a_x e^{i\int dtdx \pi\partial_t\phi}\delta(\tilde{F_a})\det\left(\frac{\partial \tilde{F_a}}{\partial \tilde{e}^b_x}\right) e^{-i L_a \epsilon^a_b\tilde{e}^b_x/\lambda}
    \label{Zconstexphipi}
\end{equation}

\noindent where we have used that at constant $e^a_x$ 

\begin{equation}
    \det\left\{\tilde{F}_a,\tilde{G}\right\} = \det\left(\frac{\partial \tilde{F_a}}{\partial \tilde{e}^b_x}\right)
\end{equation}

\noindent and $P^a = \epsilon^a_b\tilde{e}^b_x/\lambda$. The delta function that imposes $\tilde{F}_a$ fixes $\tilde{e}^a_x$ to be such that $P^a$ satisfy the $T\bar{T}$ flow equation, as explained in section \ref{classical}. This of course gives back the expression \eqref{constphasespacepartfun} for the partition function, modulo the fact that in \eqref{constphasespacepartfun} $P^a$ requires some ordering procedure whereas the expression we have just obtained $P^a$ appears to be just the phase space function obtained from solving $\tilde{F}_a=0$, this subtlety being another avatar of the ordering issues we have pointed out earlier.\\
\indent However, in order to make contact with the procedure performed in \cite{dubovsky2018tt}, some different manipulations are useful. First, the finite-dimensional delta function can be exponentiated by integrating-in a constant variable $\tilde{e}_t^a$

\begin{equation}
    \delta(\tilde{F}^a) = \int d\tilde{e}_t^a e^{-i \tilde{e}_t^aF_a} = \int d\tilde{e}_t^a e^{-i\tilde{H}_{grav}-i\tilde{H}_{0}}
\end{equation}

\noindent where $\tilde{H}_{grav}$ and $\tilde{H}_{0}$ are the zero modes of the gravity and matter Hamiltonians respectively evaluated at constant geometry $\tilde{e}^a_{\mu}$. Then the partition function can be written as 

\begin{equation}
    Z(L,R) = \int d\tilde{e}^a_x d\tilde{e}^a_t e^{-i\tilde{H}_{grav}-i L_a P^a\tilde{e}^b_x/\lambda} W\left[\tilde{e}\right] 
    \label{ZvsW}
\end{equation}

\noindent where

\begin{equation}
W\left[\tilde{e}\right] = \int D\pi D\phi e^{i\int dtdx \pi\partial_t\phi}\det\left(\frac{\partial \tilde{F_a}}{\partial \tilde{e}^b_x}\right)e^{-i\tilde{H}_0}     
\end{equation}

\noindent This last quantity $W\left[\tilde{e}\right]$ is calculated purely in terms of the matter theory (the undeformed theory) at fixed constant geometry. As we have mentioned for our starting point, we may assume that it is understood how to quantize the undeformed matter theory, in the sense that when placing the theory in a cylinder with constant geometry of size $L$ we have a basis of energy and momentum eigenstates of the undeformed theory $\ket{k}$ with eigenvalues $E_0^{(k)}(L)$ and $P_0^{(k)}(L)$ respectively (here $k$ stands for some multi-index labelling these states). Then, modulo ordering subtleties, we can write 

\begin{equation}
\label{quantumW}
W\left[\tilde{e}\right] = \sum_k \bra{k}\det\left(\frac{\partial \hat{\tilde{F_a}}}{\partial \tilde{e}^b_x}\right)e^{-i\hat{\tilde{H}}_0}\ket{k}
\end{equation}

\noindent where the hat denotes that we have promoted these functions of the phase space of the undeformed theory into operators in the Hilbert space of the undeformed theory. Both $\tilde{F}_a$ and $\tilde{H}_0$ are functions of the undeformed energy $E_0(L)$ and momentum $P_0(L)$ on a circle of size $L$, where $L$ in these expressions depends on $\tilde{e}^a_x$ via $L |e_x^2|^{1/2}$. Then naively we would want to write 

\begin{equation}
W\left[\tilde{e}\right] = \sum_k \det\left(\frac{\partial \tilde{F}^{(k)}_a}{\partial \tilde{e}^b_x}\right)e^{-i\tilde{H}^{(k)}_0}  
\label{sumW}
\end{equation}

\noindent where $\tilde{F}^{(k)}_a$ and $\tilde{H}^{(k)}_0$ denote these quantities with $E_0(L)$ and $P_0(L)$ evaluated at $E_0^{(k)}(L)$ and $P_0^{(k)}(L)$.\\
\indent What may obstruct the passage from \eqref{quantumW} to \eqref{sumW} is the fact that the expectation values summed over also include insertions of $\partial_{L}E_0(L)$ and $\partial_{L}P_0(L)$ coming from the derivatives of $\tilde{F}_a$ with respect to $\tilde{e}_x^a$. Because of the quantization of total momentum at fixed size $L$, $P_0(L)$ is necessarily of the form $P_0(L) = N/L$ for some integer $N$. Therefore $\ket{k}$ are also eigenvectors of $\partial_{L}P_0(L)$ and insertions of this operator do not introduce complications in the passage from \eqref{quantumW} to \eqref{sumW}.\\ 
\indent Let us focus on terms in \eqref{quantumW} involving insertions of $\partial_{L}E_0(L)$. Because of the structure of $\tilde{F}_a$, given by \eqref{tildeF} with $P^a = \epsilon^a_b\tilde{e}^a_x/\lambda$, these insertions can appear at the most at quadratic order in \eqref{quantumW}. However, because of the structure of the determinant it turns out that quadratic insertions of these operator vanish from \eqref{quantumW}. Terms linear in $\partial_{L}E_0(L)$ are easily taken care of by the fact that $\bra{k}\partial_{L}E_0(L)\ket{k} = \partial_LE_0^{(k)}(L)$ by standard perturbation theory arguments and thus expression \eqref{sumW} appears to be valid.\\
\indent Plugging in then \eqref{sumW} into our expression \eqref{ZvsW} for $Z(L,R)$ yields

\begin{equation}
    Z(L,R) = \sum_{k}\int d\tilde{e}^a_x d\tilde{e}^a_t e^{-i\tilde{H}_{grav}-i L_a P^a\tilde{e}^b_x/\lambda}\det\left(\frac{\partial \tilde{F}^{(k)}_a}{\partial \tilde{e}^b_x}\right)e^{-i\tilde{H}^{(k)}_0}.    
\end{equation}

\noindent This is --if not precisely the same-- of the form found for the gravitational $T\bar{T}$ partition function found in \cite{dubovsky2018tt} (see for instance equations (43) to (50) there): an integral over constant dyads on the torus and a sum over the eigenstates of the undeformed theory. However, for the integral found in \cite{dubovsky2018tt}, the fact it was one-loop exact seemed quite surprising, a conspiracy of factors coming from the non-trivial path integral measures, coming from inner products in the space of fields fixed by demanding diffeomorphism invariance.\\ 
\indent Here however this saddle-point exactness property is quite simple. Indeed, undoing in these last expression the steps going from \eqref{Zconstexphipi} to \eqref{ZvsW} by integrating over $\tilde{e}^a_t$ we get

\begin{equation}
    Z(L,R) = \sum_k\int d\tilde{e}^a_x \delta(\tilde{F}^{(k)}_a)\det\left(\frac{\partial \tilde{F}_a^{(k)}}{\partial \tilde{e}^b_x}\right) e^{-i L_a P^a(\tilde{e}_x)}
\end{equation}

\noindent Then the delta function fixes the integral to values of $\tilde{e}^a_x$ such that $P^a(\tilde{e}_x)$ coincide with the $T\bar{T}$ deformed spectrum, (as discussed in detail in section \eqref{classical}). The determinant factor gives the right measure so that 

\begin{equation}
    Z(L,R) = \sum_k e^{-iL_0 (E^{(k)}_{\lambda}(R) + R/\lambda)-iL_1P^{(k)}_{\lambda}(R) }
    \label{TTbarpartfun}
\end{equation}

\noindent without there being any extra factors. Here $E^{(k)}_{\lambda}(R)$ and $P^{(k)}_{\lambda}(R)$ are the solutions to the $T\bar{T}$ flow equations with $E^{(k)}_{0}(R)$ and $P^{(k)}_{0}(R)$ as intial ($\lambda = 0$) conditions. We have arrived thus at the $T\bar{T}$-deformed partition function.\\
\indent Let us briefly recap what we have accomplished. We started from the observation that in a particular gauge, the gauge-fixed phase space is identical to the undeformed phase space, thus we have assumed that the gauge-fixed phase space admits a canonical quantization procedure identical to that of the undeformed theory. The physical spacetime translation operators of the gravitational theory are the $P^a$, which are made into functions exclusively of this phase space after one solves the constraint equations. Since these generate physical spacetime translations, it makes sense to consider the partition function defined by \eqref{canonicalconstrpartitionfunction} over the gauge-fixed --equivalently, the undeformed-- phase space. Via an expression of this partition function as a path integral over canonical variables in the constrained phase space and a type of canonical Faddeev-Popov trick we managed to write this as a gauge-fixed form (in a particular gauge) of the covariant path-integral discussed in \cite{dubovsky2018tt}. Following a slightly different version of this procedure in reverse we have shown that this reproduces the $T\bar{T}$-deformed partition function.\\
\indent Many of the steps we have followed are dependent on how one handles ordering ambiguities that may arise, which we have mentioned at every relevant step, yet not analyzed in detail. It would seem to be difficult to do this at the level of generality about the undeformed theory we are assuming here, and perhaps this analysis is better performed in a case-by-case basis.\\ 
\indent It is interesting to point out however that in going from the gauge-fixed covariant expression \eqref{covariantPI} to the $T\bar{T}$-deformed partition function \eqref{TTbarpartfun} the only ordering issues that arose were those within the quantization of the undeformed theory, related to the operators appearing in the expectation values taken in \eqref{quantumW}. This lends some weight to the intuitive idea mentioned in section \ref{classical} that perhaps to get the fully quantum $T\bar{T}$ deformed spectrum in a sense only quantum corrections at the level of the undeformed theory need to be considered. It appears that further analysis is required from this point of view, perhaps in a case-by-case basis considering specific undeformed theories, to put these types of argument in more solid footing.  

\section{Conclusions and future directions} \label{conclusions}

In this paper we have analyzed the gravitational description of the $T\bar{T}$ deformation from the Hamiltonian point of view, where it is described as a constrained canonical system. We have found that the constraints imply non-trivial relations between the target-space observables of the theory (the physical energy and momentum) and the energy and momentum of the undeformed theory. These relations, given by \eqref{implicitTTflowE} and \eqref{implicitTTflowP} coincide with the implicit form of the solutions of the $T\bar{T}$ finite-volume flow equation \eqref{burgers} once one identifies the target space energy and momentum with the $T\bar{T}$-deformed energy and momentum. We have found that this result holds for a quite general undeformed matter content. It is interesting that these equations emerge within the gravitational description at the level of the classical equations of motion, whereas the $T\bar{T}$ deformed spectrum is supposed to hold fully quantum-mecanically. As it is well known, this already happens at the level of the lightcone-quantized critical bosonic string, and we have pointed out the similarity between the two cases.\\
\indent We have followed a similar approach to study a similar deformation, inspired by ``gravitational'' formulations of the Lorentz-breaking $J\bar{T}$ deformation proposed in the literature. Here we have found that the deformed spectrum depends on the undeformed quantities evaluated at twisted boundary conditions. Although this feature is expected if one considers $S$-matrix deformations of integrable theories, it makes comparison with previous results somewhat difficult and further work is warranted in this direction. It would also be interesting to see whether a similar analysis to that performed here can also be applied to other ``cousins'' of the $T\bar{T}$ deformation, such as the double current deformations considered in \cite{Dubovsky:2023lza}, the $T\bar{T}+\Lambda_2$ deformation discussed in \cite{Torroba:2022jrk} or the Root-$T\bar{T}$ deformation introduced in \cite{Ferko:2022cix,Borsato:2022tmu}.\\
\indent Perhaps the most interesting avenue for future work in terms of possible applications is to extend the $T\bar{T}$-gravity formulation of non-critical string theories discussed in section \ref{sec3massless} to the sector without winding, particularly for a three-dimensional target space, where the spectrum in the winding sector is known to be compatible with target-space Lorentz symmetries at the quantum level. Here instead of the finite-volume energy spectrum $E(R)$, one would be interested in the energies as function of the angular momentum $J$. A possible approach would be to consider the path integral on the Euclidean torus with boundary conditions twisted by $J$ along some cycle. A first step towards attempting this calculation is to identify how rotations act on the gravitational variables (in contrast to what happens in the Polyakov formulation, this action is non-trivial in our case), which we have done in this paper. Very likely, this path integral calculation will involve considering singular metric configurations, which adds an extra complication with respect to the non-zero winding case. In any case, it will be interesting to see whether an exact formula emerges from this procedure, as happens in the winding sector.   

%Starting from the action proposed in \cite{dubovsky2018tt}, we obtained the corresponding Hamiltonian and found that, as expected for generally covariant systems, it is a combination of constraints. The algebra of said constraints was found to be that of the canonical formulation of 1+1 dimensional string theories; the aforementioned algebra was obtained considering a fairly general matter content, since we assumed the matter Lagrangian to depend exclusively on Lorentz invariant combinations of the a real scalar field and its first derivatives. We wrote down the mode zero of the constraint equations in terms of Dirac observables and observed that they are analogous to (the implicit solution of) the $T\Bar{T}$ energy-flow equation. Similar results were obtained for the case of a complex scalar field and the so called $J\Bar{T}$ deformations. We then considered the case of a free, massless scalarfield and upon extending the phase space to include the matter fields, we found out that {\color{red} ?},

\section*{Acknowledgements}
We thank Miguel Campiglia, Rodolfo Gambini and Sergei Dubovsky for many helpful discussions regarding this work. GHC acknowledges the support of CSIC I+D grant number 583. This work is also supported by PEDECIBA.

\appendix

\section{Constrained Hamiltonian Systems}
\label{appconstrainedsystems}
In this section, we will introduce the most general aspects of constrained Hamiltonian systems (which include gauge systems). Part of the importance of having an appropriate Hamiltonian formulation for constrained systems lies in the need to develop a valid canonical quantization procedure.

It can be shown (see \cite{henneaux}) that the total Hamiltonian in these cases is of the form
\begin{equation}
    H_T=H+u^mc_m
    \label{totalham}
\end{equation}
where the  $u^m$ are Lagrange multipliers of the constraints $c_m$,  and $H=\dot{q}^n p_n-L$.

In order to appreciate the relation between quantities that appear in the Hamiltonian formalism, a distinction between first and second class functions is usually made. This relies only on the fundamental structure of the Hamiltonian theory, the Poisson bracket. Any quantity that has a weakly null bracket with all the constraints is called first class. Quantities that do not have this property are called second class. The Total Hamiltonian \eqref{totalham} is expressed as the sum of a first-class Hamiltonian plus a linear combination of first-class constraints.

All the first-class constraints generate gauge transformations that are independent if and only if the constraints are irreducible.The transformation generated by a first-class constraint preserves all first and second-class constraints, so it maps an allowed state onto an allowed state. Also, the Poisson brackets of any two first-class constraints is a first-class constraint and generates a gauge transformation.

On the other hand, second-class constraints are present if the matrix $C_{j j'}=[c_j,c_{j'}]$ does not vanish on the constraint surface. This kind of constraints do not generate any transformation of physical significance, including gauge transformations, because the transformation generated by a second-class constraint does not preserve all the constraints of the system, so it maps an allowed state onto a nonallowed state.
To treat second-class constraints we must define the Dirac bracket:
\begin{equation}
    [F,G]^*=[F,G]-[F,c_i]C^{ij}[c_j,G]
\end{equation}

If the theory is formulated in terms of the Dirac brackets, the second-class constraints become strong identities expressing canonical variables in terms of others, and in some cases they can be used to eliminate some degrees of freedom.

\section{Hamiltonian for scalar fields with one derivative interactions in gravitational background}
\label{appscalarham}

In order to obtain the Hamiltonian for the scalar matter theory in the gravitational background we must first find the canonical momentum $\pi$ of the matter field in terms of the time derivatives of $\phi$. This is accomplished via the defining relation 

\begin{equation}
    \pi = \sqrt{-g}\partial_u K(u,\phi)\frac{\partial u}{\partial \partial_t \phi}
\end{equation}

\noindent Using that $u$ can be written as

\begin{equation}
    u = r\left(\sqrt{|g^{00}|}\partial_t \phi + r \frac{g^{01}}{\sqrt{|g^{00}|}}\partial_x \phi\right)^2 + \frac{1}{g^{00}g}(\partial_x\phi)^2 
\end{equation}

\noindent where $r\equiv \sign(g^{00})$. The equation to invert $\partial_t\phi$ in terms of $\pi$ is thus of the form

\begin{equation}
\sqrt{|g^{00}|}\partial_t \phi + r \frac{g^{01}}{\sqrt{|g^{00}|}}\partial_x \phi = f\left(\frac{\pi}{\sqrt{|g^{00}g|}},\frac{(\partial_x\phi)^2}{g^{00}g},\phi\right)
\end{equation}

\noindent for some function $f$ of the variables made explicit above, the specific details of which depend on the particular form of $K$. Using this we get that the Hamiltonian can be written as

\begin{align}
    H_{mat} &\equiv \int dx\pi \partial_t \phi - \int dx \sqrt{-g}K(u,\phi)\nonumber\\
    &=  \int dx \left(- \frac{g^{01}}{g^{00}}\pi \partial_x \phi + \frac{\pi}{\sqrt{|g^{00}|}}f - \sqrt{-g}K\left(f^2+\frac{(\partial_x\phi)^2}{g^{00}g},\phi\right)\right)\nonumber\\
    &= \int dx \left(- \frac{g^{01}}{g^{00}}\pi \partial_x \phi + \sqrt{-g}\tilde{f}\left(\frac{\pi}{\sqrt{|g^{00}g|}},\frac{(\partial_x\phi)^2}{g^{00}g},\phi\right)\right)
\end{align}

\noindent from which \eqref{matterhamiltonian} immediately follows. Notice also that by evaluating the previous expression in flat space $g_{\mu\nu} = \eta_{\mu\nu}$  it is clear that $\tilde{f}$ is the flat-space matter Hamiltonian density, so that by a change of variables it is seen that the interpretation of \eqref{undeformedenergyL} given in the main text follows. 

\section{Constraint algebra}\label{appConstAlg}

In this section we are going to show some details of the calculation of the constraint algebra and compute the Poisson brackets  \eqref{PoissonCC}.
Defining the gravitational part of \eqref{constraintC}  as

\begin{equation}
    \mathcal{C}_{grav}[N] = \int N ~ e_{x}^{a}\eta_{ab}(e_{x}^{b}+\partial_{x}X^{b}),
\end{equation}
and the material part as
\begin{equation}
    \mathcal{C}_{mat}[N] = \int N\mathcal{H}_{0}(x)
\end{equation}
we have
\begin{align}
\nonumber
\left\lbrace \mathcal{C}[N],\mathcal{C}[M]\right\rbrace &=\left\lbrace \mathcal{C}_{grav}[N],\mathcal{C}_{grav}[M] \right\rbrace +\left\lbrace \mathcal{C}_{mat}[N],\mathcal{C}_{mat}[M]\right\rbrace 
\\
\label{AppCC2terms}.
&+ \left\lbrace \mathcal{C}_{grav}[N] , \mathcal{C}_{mat} [M] \right\rbrace + \left\lbrace \mathcal{C}_{mat}[N] , \mathcal{C}_{grav} [M] \right\rbrace
\end{align}

The first term in  \eqref{AppCC2terms} corresponds to the Poisson brackets between the gravitational part. This is
\begin{align}
\label{PB_HGrav_HGrav}
    \left\lbrace \int \mathrm{d}x N ~ e_{x}^{a}\eta_{ab}(e_{x}^{b}+\partial_{x}X^{b}) , \int \mathrm{d}x' M ~ e_{x}^{a'}\eta_{a'b'}(e_{x}^{b'}+\partial_{x'} X^{b'}) \right\rbrace
\end{align}
Before continuing, we first note that the only terms that contribute to the final result are those involving the Poisson brackets $\left\lbrace \partial_x X^b , e^{a'}_x \right\rbrace$ and $\left\lbrace e^a_x , \partial_x X^{b'} \right\rbrace$, while the other two terms cancel due to the anti symmetry of the Levi-Civita (which appears when taking the Poisson bracket of the canonical variables). With this in mind, we can write \eqref{PB_HGrav_HGrav} as
\begin{align}
   \int N(x) M(x') e^x_a(x) \eta_{a b} \left\lbrace \partial_x X^b (x) , e^{a'}_x (x') \right\rbrace \eta_{a'b'} (e^{b'}_x (x')+ \partial_{x'} X^{b'}(x'))
   \\
\nonumber
+\int N(x) M(x') \left\lbrace e^{a}_x (x) , \partial_{x'} X^{b'} (x') \right\rbrace \eta_{a b} (e^b_x (x) + \partial_x X^b (x)) e^{a'}_x (x') \eta_{a'b'}.
\end{align}
It is a simple task to verify that only each second term in the curved parenthesis contributes to the final result. We are then left with
{\allowdisplaybreaks
\begin{align}
  & \int N(x) M(x') e^x_a(x) \eta_{a b} \left\lbrace \partial_x X^b (x) , e^{a'}_x (x') \right\rbrace \eta_{a'b'} \partial_{x'} X^{b'}(x')
   \\
\nonumber
& +\int N(x) M(x') \left\lbrace e^{a}_x (x) , \partial_{x'} X^{b'} (x') \right\rbrace \eta_{a b} \partial_x X^b (x) e^{a'}_x (x') \eta_{a'b'}
\\
\nonumber
& = \int N(x) M(x') e^x_a(x) \partial_{x'} X^{b'}(x') \eta_{a b} \varepsilon^{ba'} \eta_{a'b'} \partial_x \delta(x-x')  
   \\
\nonumber
& +\int N(x) M(x') e^{a'}_x (x') \partial_{x} X^b (x)  \eta_{a b} \varepsilon^{ab'} \eta_{a'b'}   \partial_{x'} \delta(x-x')
\\
\nonumber
&=\int N(x) M(x') e^x_a(x) \varepsilon_{ab} \partial_{x'} X^{b}(x') \partial_x \delta(x-x') 
\\
\nonumber
&+ \int N(x) M(x') e^x_a(x') \varepsilon_{ab} \partial_x X^{b}(x) \partial_{x'} \delta(x-x').
\\
\nonumber
\end{align}
Integrating over the Dirac delta, and noticing that the only non vanishing terms are those involving the derivative of the Lagrange multipliers, we finally get
\begin{align}
    \left\lbrace \mathcal{C}_{grav} [N] , \mathcal{C}_{grav}[M] \right\rbrace &= \int \mathrm{d}x (N \partial_x M - M \partial_x N) e^x_a \varepsilon_{a b} \partial_x X^b  
    \\
    \nonumber
    &= \mathcal{C}_{x,grav} [N \partial_x M - M \partial_x N],
\end{align}
where $C_{x,grav}$ stands for the gravitational part of the diffeomorphism constraint.\\
\indent We now proceed to calculate the second term in \eqref{AppCC2terms}, which will be the most complicated because of the degree of generality we are assuming for the matter action. We start by remembering that the matter Hamiltonian density is obtained after performing a Legendre transform
\begin{align}
    \mathcal{H}_{mat} = \pi \partial_t \phi - \sqrt{-g} K,
\end{align}
where $\sqrt{-g} K$ is the Lagrangian density and we have assumed it to be of the form
\begin{align}
    K=K(\phi,u), ~~ u=\partial_\mu \phi ~ \partial^\mu \phi = g^{\mu\nu} \partial_\mu \phi ~ \partial_\nu \phi,
\end{align}
so that $K$ depends exclusively on Lorentz invariant combinations of the field and its derivatives up to second order. $\pi$ is the conjugate momentum associated to the field $\phi$ and is defined as
\begin{align}
\label{material_momentum}
    \pi = \sqrt{-g} \frac{\partial K}{\partial (\partial_t \phi)}
\end{align}
Let us now perform some preliminary calculations:
\begin{align}
    \left\lbrace \pi(x) \partial_t \phi (x), \phi (x') \right\rbrace &= - \partial_t \phi (x) \delta(x-x') - \pi(x) \frac{\delta \partial_t \phi(x)}{\delta \pi (x')}
    \\
    \nonumber
    \left\lbrace \pi(x) \partial_t \phi (x), \pi (x') \right\rbrace &= \pi(x) \frac{\delta \partial_t \phi(x)}{\delta \phi(x')}
    \\
    \nonumber
    \left\lbrace \sqrt{-g}(x) K(x) , \phi(x') \right\rbrace &= - \sqrt{-g} \pi(x) \frac{\delta \partial_t \phi(x)}{\delta \pi (y)}
    \\
    \nonumber
   \left\lbrace \sqrt{-g}(x) K(x) , \pi (x') \right\rbrace &= \sqrt{-g} \left. \frac{\delta K(x)}{\delta \phi(x')}\right|_{\partial_t \phi} + \pi(x) \frac{\delta \partial_t \phi (x)}{\delta \phi(x')},
\end{align}
where we used the shorthand notation $ K(x)=K(\phi(x),u(x)) $ and $\partial_t \phi(x)$ is determined by inverting equation \eqref{material_momentum}, the vertical bar on the first term on the RHS denoting that the functional derivative is taken at constant $\partial_t \phi(x)$. With these results we have
\begin{align}
\label{basic_brackets}
    \left\lbrace \mathcal{H}_{mat}(x) , \phi(x') \right\rbrace = - \partial_t \phi(x) \delta(x-x')
    \\
    \nonumber
    \left\lbrace \mathcal{H}_{mat}(x) , \pi(x') \right\rbrace = - \sqrt{-g} \left. \frac{\delta K}{\delta \phi (x')} \right|_{\partial_t \phi}.
\end{align}
We now come to the key trick that is used to perform the calculation. Notice that the matter part of the $\mathcal{C}[N]$ constraints, that is $\mathcal{C}_{mat}[N]$, can be obtained from $\int dx \mathcal{H}_{mat}(x)$ by the substitution 
\begin{align}
    e_t^a \rightarrow N e_x^b \varepsilon^a_b. 
\end{align}
\noindent However, due to the arguments in appendix \ref{appscalarham}, $\mathcal{H}_{mat}$ is linear in $e_t^a$ and therefore

\begin{equation}
    \mathcal{C}_{mat}[N] = \int dx \mathcal{H}_{mat}\rvert_{e_t^a \rightarrow N e_x^b \varepsilon^a_b} = \int dx N(x)\mathcal{H}_{mat}\rvert_{e_t^a \rightarrow e_x^b \varepsilon^a_b}.
\end{equation}

\noindent Thus, in order to calculate $\left\{\mathcal{C}_{matter}[N],\mathcal{C}_{matter}[M]\right\}$ we start by calculating
\begin{align}
\nonumber
    &\left\lbrace \int \mathrm{d}x N(x)\mathcal{H}_{mat} (x)  , \int \mathrm{d}yM(y) \mathcal{H}_{mat} (y) \right\rbrace = \int \mathrm{d}x \mathrm{d}y N(x)M(y) \left\lbrace \mathcal{H}_{mat} (x),\mathcal{H}_{mat} (y) \right\rbrace 
    \\
    &= \int \mathrm{d}x \mathrm{d}y N(x)M(y)\mathrm{d}z \left\lbrace \frac{\delta \mathcal{H}_{mat} (x)}{\delta \phi(z)} \frac{\delta \mathcal{H}_{mat} (y)}{\delta \pi(z)} - \frac{\delta \mathcal{H}_{mat} (x)}{\delta \pi(z)} \frac{\delta \mathcal{H}_{mat} (y)}{\delta \phi(z)} \right\rbrace
\end{align}
and we will later on evaluate this at $e_t^a \rightarrow e_x^b \varepsilon^a_b$. Using now the results \eqref{basic_brackets}, we get
\begin{align}
\label{PB_Cmat_Cmat}
&\left\lbrace \int \mathrm{d}x N(x)\mathcal{H}_{mat} (x)  , \int \mathrm{d}y N(y)\mathcal{H}_{mat} (y) \right\rbrace\nonumber\\ &= \int N(x)N(y)\left[ \delta(x-z) \partial_t \phi(x) \sqrt{-g}(y) \frac{\delta K(y)}{\delta \phi(z)} - (x \leftrightarrow y) \right],
\end{align}
where we have now omitted the measures of integration. The functional derivative appearing in the previous equation can be calculated as
\begin{align}
\frac{\delta K(x)}{\delta \phi(y)} = \partial_\phi K(x) \delta(x-y) + \partial_u K(x) \left. \frac{\delta u(x)}{\delta \phi (y)} \right|_{\partial_t \phi}
\end{align}
When inserting this expression into \eqref{PB_Cmat_Cmat}, we can see that the contribution proportional to the first term vanishes due to the antisymmetrization, so we are left with
\begin{align}
\label{PB_Cmat_Cmat_2}
&\left\lbrace \int \mathrm{d}x N(x)\mathcal{H}_{mat} (x)  , \int \mathrm{d}y M(y)\mathcal{H}_{mat} (y) \right\rbrace \nonumber\\&= \int N(x)M(y)\left[ \delta(x-z) \partial_t \phi(x) \sqrt{-g}(y) \partial_u K(y) \left. \frac{\delta u(y)}{\delta \phi(z)} \right|_{\partial_t \phi} - (x \leftrightarrow y) \right].
\end{align}
Since $u= g^{\mu\nu} \partial_\mu \phi \partial_\nu \phi$, we have
\begin{align}
\label{derivative_u_phi}
    \left. \frac{\delta u(y)}{\delta \phi(z)} \right|_{\partial_t \phi} = 2 g^{01} \partial_t \phi(y) \partial_y \delta(y-z) + 2 g^{11} \partial_y \phi(y) \partial_y \delta(y-z) 
\end{align}
%Substituting this expression in \eqref{PB_Cmat_Cmat_2} and integrating out the delta, we are left only with those terms which involve the derivative of %the Lagrange multipliers:
%\begin{align}
%\label{PB_Cmat_Cmat_3}
%\left\lbrace \mathcal{C}_{mat} [N] , \mathcal{C}_{mat} [M] \right\rbrace = \int (N \partial_x M - M \partial_x N ) \partial_t \phi \sqrt{-g} \partial_u K %~ 2(g^{01} \partial_t \phi + g^{11} \partial_x \phi)
%\end{align}
From the expression for the metric in terms of the dyad $g_{\mu\nu}= \eta_{ab} e_\mu^a e_\nu^b$, we then have
\begin{align}
    g^{\mu\nu}=\frac{1}{g}\begin{pmatrix}
        e_x^a e_{x a} & -e_t^a e_{x a}
        \\
        -e_t^a e_{x a} & e_t^a e_{t a}
    \end{pmatrix}
\end{align}
We now evaluate at $e_t^a= e^b_x \varepsilon^a_b$. With this choice of $e_t^a$, \eqref{derivative_u_phi} becomes
\begin{align}
\label{derivative_u_phi2}
    \left. \frac{\delta u(y)}{\delta \phi(z)} \right|_{\partial_t \phi} = 2 g^{11} \partial_y \phi(y) \partial_y \delta(y-z) = 2 g^{00} \partial_y \phi(y) \partial_y \delta(y-z).
\end{align}
Then, we have
\begin{align}
\label{PB_Cmat_Cmat_3}
\left. \left\lbrace \int \mathrm{d}x N(x) \mathcal{H}_{mat} (x)  , \int \mathrm{d}y M(y) \mathcal{H}_{mat} (y) \right\rbrace \right|_{e_t^a= e_x^b \varepsilon^a_b}= 
\\
\nonumber
\int N(x) M(y) \left[ \delta(x-z) \partial_t \phi(x) \sqrt{-g}(y) \partial_u K(y) \left. \frac{\delta u(y)}{\delta \phi(z)} \right|_{\partial_t \phi} - (x \leftrightarrow y) \right]
\\
\nonumber
\int N(x) M(y) \left[ \delta(x-z) \partial_t \phi(x) \sqrt{-g}(y) \partial_u K(y) 2 g^{00} \partial_y \phi(y) \partial_y \delta(y-z) - (x \leftrightarrow y) \right]
\end{align}
Integrating over the Dirac delta (keeping in mind that only the terms involving the derivatives of $N(x)$ or $M(y)$ will survive due to the antisymmetrizacion) and then integrating on $z$ and on one of the other spatial variables, we obtain
\begin{align}
\label{PB_Cmat_Cmat_3_2}
\left. \left\lbrace \int \mathrm{d}x N(x) \mathcal{H}_{mat} (x)  , \int \mathrm{d}y M(y) \mathcal{H}_{mat} (y) \right\rbrace \right|_{e_t^a= e_x^b \varepsilon^a_b}= 
\\
\nonumber
\int (N \partial_x M - M \partial_x N ) \left[ \sqrt{-g} ~ \partial_u K ~ \partial_t \phi ~ 2 ~ g^{00} \partial_x \phi \right].
\end{align}
Let us now recall the definition of $\pi$:
\begin{align}
\label{material_momentum_2}
    \pi = \sqrt{-g} \frac{\partial K}{\partial (\partial_t \phi)} = \sqrt{-g} ~ \partial_u K \frac{\partial u}{\partial (\partial_t \phi)} = \sqrt{-g} \partial_u K (2 g^{00} \partial_t \phi + 2 g^{01} \partial_x \phi).
\end{align}
With the choice of $e_t^a = e_x^b \varepsilon^a_b$, we have $g^{01}=0$, so \eqref{PB_Cmat_Cmat_3_2} becomes
\begin{align}
\label{PB_Cmat_Cmat_3_3}
\left. \left\lbrace \int \mathrm{d}x N(x) \mathcal{H}_{mat} (x)  , \int \mathrm{d}y M(y) \mathcal{H}_{mat} (y) \right\rbrace \right|_{e_t^a= e_x^b \varepsilon^a_b} &= \int (N \partial_x M - M \partial_x N ) \pi \partial_x \phi
\\
\nonumber
&= \mathcal{C}_{x,mat} [N \partial_x M - M \partial_x N]
\end{align}
where in the last equality we have identified the matter contribution to the $\mathcal{C}_{x}$ constraint.\\
\indent Finally, it is possible to verify that the last two terms in \eqref{AppCC2terms} vanish. In order to see this, it is enough to note that
\begin{align}
    \left\lbrace \mathcal{C}_{grav} [N] , \mathcal{C}_{mat} [M] \right\rbrace\nonumber
\end{align}
will be proportional to
\begin{align}
\label{appCaux2}
    \left\lbrace e_x^a \varepsilon^b_a \varepsilon_{bc} \partial_x X^c (x) , \left. \left( \pi \partial_t \phi - \sqrt{-g} K \right)(y) \right|_{e_t^a=e_x^b \varepsilon_b^a} \right\rbrace
\end{align}
Since the second term depends on the gravitational variables only through the dyad, we can write the previous term as
\begin{align}
     e_x^a \varepsilon^b_a \varepsilon_{bc} \left\lbrace \partial_x X^c (x) , \left. \left( \pi \partial_t \phi - \sqrt{-g} K \right)(y) \right|_{e_t^a=e_x^b \varepsilon_b^a} \right\rbrace = \partial_x \delta(x-y) e_x^a \varepsilon^b_a   \frac{ \delta \left.\left( \pi \partial_t \phi - \sqrt{-g} K \right) \right|_{e_t^a=e_x^b \varepsilon_b^a} }{\delta e^b_x}    
\end{align}
The function $\pi \partial_t - \sqrt{-g} K$ depends on $e^b_x$ both explicitly and through $\partial_t \phi$, therefore
\begin{align}
    \frac{ \delta \left. \left( \pi \partial_t \phi - \sqrt{-g} K \right)  \right|_{e_t^a=e_x^b \varepsilon_b^a} }{\delta e^b_x} &= \left. \left( \frac{ \partial \left. \left( \pi \partial_t \phi - \sqrt{-g} K \right)  \right|_{e_t^a=e_x^b \varepsilon_b^a} }{\partial e^b_x}  \right)\right|_{\partial_t \phi }
   \\
   \nonumber
   &+ \left. \left( \frac{ \partial \left. \left( \pi \partial_t \phi - \sqrt{-g} K \right) \right|_{e_t^a=e_x^b \varepsilon_b^a} }{\partial \left(\partial_t \phi \right)} \right)\right|_{\pi} \frac{\partial \left. (\partial_t \phi)  \right|_{e_t^a=e_x^b \varepsilon_b^a}}{\partial e^b_x}  
\end{align}
The second term vanishes when taking the derivative with respect to $\partial_t \phi$ while holding $\pi$ fixed by virtue of the properties of the Legendre transform. We are then left with
\begin{align}
\label{appCaux}
    \left. \left( \frac{ \partial \left. \left( \pi \partial_t \phi - \sqrt{-g} K \right)  \right|_{e_t^a=e_x^b \varepsilon_b^a} }{\partial e^b_x}  \right)\right|_{\partial_t \phi }
\end{align}
The first term in \eqref{appCaux} vanishes since the derivative is taken while holding $\partial_t \phi$ fixed. Moreover, keeping in mind that we are also evaluating on $e^a_t = e^b_x \varepsilon^a_b$, using the chain rule, we have that
\begin{align}
    \left. \left( \frac{ \partial \left. \left( - \sqrt{-g} K \right)  \right|_{e_t^a=e_x^b \varepsilon_b^a} }{\partial e^b_x}  \right)\right|_{\partial_t \phi } \propto \frac{\partial (\eta_{ac} e_x^a e_x^c)}{\partial e^b_x} 
\end{align}
When holding $\partial_t \phi$ fixed and evaluating on $e^a_t = e^b_x \varepsilon^a_b$, the term $- \sqrt{-g} K$ can be regarded as a function of $e_x^a e_x^b \eta_{ab}$. This allows us to obtain the following result for \eqref{appCaux2}
\begin{align}
    \left\lbrace e_x^a \varepsilon^b_a \varepsilon_{bc} \partial_x X^c (x) , \left. \left( \pi \partial_t \phi - \sqrt{-g} K \right)(y) \right|_{e_t^a=e_x^b \varepsilon_b^a} \right\rbrace &\propto e^a_x \varepsilon^b_a \frac{\partial (\eta_{cd}) e^c_x e^d_x}{\partial e^b_x}
    \\
    \nonumber
    &\propto e^a_x \varepsilon^b_a \eta_{b c} e^c_x = e^a_x e^b_x \varepsilon_{ab} = 0
\end{align}
So that the last two terms in \eqref{AppCC2terms} vanish, giving us the desired result
\begin{align}
    \left\lbrace \mathcal{C} [N] , \mathcal{C}[M] \right\rbrace = \mathcal{C}_{x} [N\partial_x M - M\partial_x N]
\end{align}

\bibliography{grefs}

\end{document}